\documentclass[twocolumn,aps,prb,superscriptaddress]{revtex4-1}
\usepackage[latin9]{inputenc}
\usepackage{amsmath}
\usepackage{amssymb}
\usepackage{graphicx}
\usepackage{pslatex}
\usepackage[colorlinks=true, urlcolor=blue, linkcolor=red, citecolor=blue, pdftex]{hyperref}

\makeatletter

\providecommand{\tabularnewline}{\\}
\newcommand{\lyxdot}{.}

\@ifundefined{textcolor}{}
{%
 \definecolor{BLACK}{gray}{0}
 \definecolor{WHITE}{gray}{1}
 \definecolor{RED}{rgb}{1,0,0}
 \definecolor{GREEN}{rgb}{0,1,0}
 \definecolor{BLUE}{rgb}{0,0,1}
 \definecolor{CYAN}{cmyk}{1,0,0,0}
 \definecolor{MAGENTA}{cmyk}{0,1,0,0}
 \definecolor{YELLOW}{cmyk}{0,0,1,0}
}

\usepackage{amsfonts}
\usepackage{color}

\makeatother

\begin{document}

\title{Analytical and numerical studies of disordered spin-1 Heisenberg chains with aperiodic couplings}

\author{H. L. Casa Grande}
\affiliation{Instituto de F\'{\i}sica, Universidade de S\~ao Paulo, Caixa Postal 66318,
05314-970, S\~ao Paulo, SP, Brazil}
\affiliation{Laboratoire de Physique Th\'eorique, IRSAMC, Universit\'e de Toulouse, CNRS, 31062 Toulouse, France}
\author{N. Laflorencie}
\affiliation{Laboratoire de Physique Th\'eorique, IRSAMC, Universit\'e de Toulouse, CNRS, 31062 Toulouse, France}
\author{F. Alet}
\affiliation{Laboratoire de Physique Th\'eorique, IRSAMC, Universit\'e de Toulouse, CNRS, 31062 Toulouse, France}
\author{A. P. Vieira}
\affiliation{Instituto de F\'{\i}sica, Universidade de S\~ao Paulo, Caixa Postal 66318,
05314-970, S\~ao Paulo, SP, Brazil}

\begin{abstract}
We investigate the low-temperature properties of the one-dimensional 
spin-1 Heisenberg model with geometric fluctuations induced by aperiodic
but deterministic coupling distributions, involving two parameters.
We focus on two aperiodic sequences, the Fibonacci sequence and the
6-3 sequence. Our goal is to understand how these geometric fluctuations
modify the physics of the (gapped) Haldane phase, which corresponds
to the ground state of the uniform spin-1 chain. We make use of different
adaptations of the strong-disorder renormalization-group (SDRG) scheme
of Ma, Dasgupta and Hu, widely employed in the study of random spin
chains, supplemented by quantum Monte Carlo and density-matrix renormalization-group
numerical calculations, to study the nature of the ground state as
the coupling modulation is increased. We find no phase transition
for the Fibonacci chain, while we show that the 6-3 chain exhibits
a phase transition to a gapless, aperiodicity-dominated phase similar
to the one found for the aperiodic spin-1/2 XXZ chain. Contrary to
what is verified for random spin-1 chains, we show that different
adaptations of the SDRG scheme may lead to different qualitative conclusions
about the nature of the ground state in the presence of aperiodic coupling
modulations.
\end{abstract}
\maketitle

\section{Introduction}

Quantum spin chains represent a suitable laboratory for the study
of the combined effects, on many-body systems, of quantum fluctuations
and broken translation symmetry, represented for instance by the presence
of an inhomogeneous coupling distribution. A paradigmatic model in
this context is the Heisenberg chain, described by the Hamiltonian
\begin{equation}
H=\sum_{j=1}^{L}J_{j}\mathbf{S}_{j}\cdot\mathbf{S}_{j+1},
\label{eq:H}
\end{equation}
in which the constants $J_{j}>0$ are antiferromagnetic couplings
between the spin-$S$ operators located at contiguous sites. 

Even in the uniform limit ($J_{j}\equiv J$), the model exhibits a
variety of physical behavior, strongly dependent on the integer or
half-integer character of $S$. According to a widely accepted 
conjecture by Haldane,\cite{haldane83,affleck89} chains with half-integer $S$
have a gapless energy spectrum, while the ground state of chains with
integer $S$ is separated from the first excited states by a finite energy
gap. The most notable effects are seen in the extreme quantum limit
of small values of $S$, in which the two classes of systems are represented
by $S=\frac{1}{2}$ and $S=1$. In this last case ($S=1$), the ground
state --- the so-called Haldane phase --- which can be well approximated
by a valence-bond-solid state,\cite{affleck87} exhibits a hidden topological
order, revealed by a string order parameter,\cite{dennijs}
and the boundaries of open finite chains of length $L$ harbor spin-$\frac{1}{2}$
degrees of freedom.

Whether the low-energy spectrum is gapless or gapped governs not only
the low-temperature thermodynamic behavior, but also affects the stability
of the uniform ground state towards the breaking of translation symmetry.
In the simple case of dimerization (the introduction of alternating
couplings $J_{\mathrm{odd}}$ and $J_{\mathrm{even}}$ along the chain),
the spin-$\frac{1}{2}$ chain becomes gapped even in the presence
of an infinitesimal difference between $J_{\mathrm{odd}}$ and $J_{\mathrm{even}}$,
while the Haldane phase is protected by the finite gap.

Disorder effects, represented by random uncorrelated couplings, are
even more pronounced. For the spin-$\frac{1}{2}$ chain, much information
on the effects of random couplings can be obtained by using the strong-disorder 
renormalization-group (SDRG) scheme introduced by Ma, Dasgupta, and Hu.\cite{ma79,dasgupta80}
The basic idea is to eliminate high-energy degrees of freedom
by identifying strongly coupled spin pairs along the chain, which contribute very little 
to magnetic properties at low temperatures and therefore can be decimated away, giving 
rise to weak effective couplings between the remaining neighboring spins. A
number of studies performed during the last two decades\cite{doty92,fisher94,henelius98,Laflo04,hoyos07}
showed that, in the presence of any finite disorder, the ground state
turns into a random-singlet phase, which can be pictured as a collection
of widely separated spin pairs coupled in singlet states. This is
a consequence of the fact that, in the renormalization-group language,
disorder induces a flow of the probability distribution of effective
couplings towards an infinite-randomness fixed point, in which, at a
given energy scale, there are only a few strong effective couplings,
which give rise to the singlet pairs, while the vast majority of the
remaining couplings are much weaker. In this random-singlet phase,
physical properties are quite distinct from the ones in the uniform
chain, being characterized by an activated dynamics, in which energy
and length scales are not related by a power law, but by a stretched
exponential form. Furthermore, ground-state spin-spin correlations
are dominated by the rare singlet pairs, leading to a striking distinction
between average and typical behaviors.\cite{fisher94}

The picture for random spin-$1$ chains looks even richer. Investigations
based on extensions of the SDRG scheme,\cite{damle02,saguia02,hyman97,
hyman96phd,monthus97,monthus98}
combined with numerical studies,
\cite{bergkvist02,lajko05,nishiyama1998,hida1999,yang2000,todo2000} 
point
to the stability of the Haldane phase towards sufficiently weak disorder;
intermediate disorder seems to lead to a gapless Haldane phase, characterized
by a finite string order parameter and exhibiting nonuniversal effects 
associated with Griffiths singularities;\cite{griffiths69}
and sufficiently strong disorder induces a random-singlet phase.

For the spin-$\frac{1}{2}$ chain, effects partially similar to
those produced by randomness are induced by the presence of aperiodic
but deterministic couplings. This kind of aperiodicity is suggested
by analogy with quasicrystals,\cite{schechtman84} structures exhibiting
symmetries forbidden by traditional crystallography and corresponding
to projections of higher-dimensional Bravais lattices onto low-dimensional
subspaces. Aperiodic couplings can be produced by letter sequences
generated by substitution rules such as the one associated with the
Fibonacci sequence, $a\rightarrow ab$, $b\rightarrow a$. The iteration
of the rule leads to a sequence $abaababa\ldots$, in which there
is no characteristic period. Associating different letters with different
coupling values $J_{a}$ and $J_{b}$, an aperiodic chain
is built. Distinct aperiodic sequences give rise to different geometric
fluctuations, gauged by a wandering exponent $\omega$ associated with
the power-law describing the growth of suitably defined coupling fluctuations as the
chain length increases. The case $\omega=\frac{1}{2}$ emulates the
fluctuations induced by random uncorrelated couplings.

An adaptation of the SDRG method to the aperiodic spin-$\frac{1}{2}$
XXZ chain,\cite{vieira05a,vieira05b} a particular case of
which is the Heisenberg chain, revealed that low-temperature thermodynamic
properties and the nature of the ground state are deeply changed by
aperiodicity generated by sequences for which $\omega\geq0$, and
behavior reminiscent of that characterizing the random-singlet phase
can be observed. Notably, there is in general a stretched exponential
relation between energy and length scales, and a clear distinction
between average and typical behavior of spin-spin correlation functions,
in this case associated with the existence of characteristic lengths
emerging from the combination of aperiodicity and quantum fluctuations.
Furthermore, and in contrast to the random-singlet phase, correlations
may exhibit an ultrametric structure related to the inflation symmetry
inherent to aperiodic sequences.

In this paper, we investigate the effects of aperiodic couplings on the
low-temperature properties of quantum spin-$1$ Heisenberg chains. As in the 
case of random uncorrelated couplings, we expect that the Haldane phase is stable 
towards the introduction of weak aperiodic modulation (as measured by a coupling 
ratio $r=J_b/J_a\simeq 1$), but that strong modulation ($r\rightarrow 0$ or 
$r\rightarrow\infty$) may lead to an aperiodicity-dominated gapless 
phase, quite similar to the one observed for the corresponding spin-$1\over 2$ 
chain. We obtain analytical results in the case of strong modulation by 
using different adaptations of the SDRG scheme. Analytical results are compared to numerical simulations obtained using quantum Monte Carlo (QMC) and density matrix renormalization group (DMRG) methods.

The first adaptation --- or approach --- of SDRG is valid only in the limit of very strong modulation, 
and corresponds to the immediate extension to spin-1 particles of the SDRG approach 
of Refs. \onlinecite{ma79} and \onlinecite{dasgupta80}. This involves identifying the most strongly 
connected spin cluster in the chain, and calculating effective couplings between spins neighboring the
cluster by assuming that the cluster is locked in its ground state. 
In the simplest case in which such cluster is a 
spin pair, the ground state is a singlet, and the excited states consist of a triplet and a quintuplet, 
all of which are only assumed to contribute to the effective couplings via virtual excitations. 
Since this fails for intermediate disorder,\cite{hyman97} a number of alternative adaptations have 
been proposed.\cite{hyman96phd, monthus97, monthus98, saguia02} One of these --- the second approach in 
the present paper --- consists in ignoring 
only the highest local excitations, usually introducing effective spins in the process, 
and calculating effective couplings so that local gaps are preserved. In case the
most strongly correlated cluster is a spin pair, this amounts to replacing the pair of $S=1$ spins 
by a pair of $S={1\over 2}$ spins, connected by a bond identical to the original one. This process is
known \emph{not} to preserve all matrix elements in the subspace of local states 
kept,\cite{hyman96phd, monthus98} a problem that can be corrected at the expense of introducing 
nonfrustrating ferromagnetic next-nearest neighbor couplings --- the third approach. 

For random uncorrelated couplings, the second and third approaches described above are expected to 
lead to the same qualitative results. However, we show here that this is not necessarily the case
in the presence of aperiodic but deterministic couplings. We argue that the qualitative equivalence
between the second and third approaches is to be expected only when geometric fluctuations, as measured
by the wandering exponent $\omega$, are sufficiently strong. 

% We consider the spin 1 Heisenberg chain with coupling modulated according
% to two aperiodic sequences. This modulation will induce some geometric
% fluctuation effects in the system and break the translational invariance.
% The first aperiodic sequence used is the Fibonacci sequence defined
% by generating a sequence as $abaababaabaababaa\dots$. The distribution
% of couplings is obtained by the association $a\rightarrow J_{a}$
% and $b\rightarrow J_{b}$.

% In the spin 1/2 chain contest this sequence is associated with marginally
% relevant geometric fluctuations, with a wandering exponent $\omega=0$,
% and it drives the system towards a phase transition to an aperiodic
% singlet phase \cite{vieira05a,vieira05b,hermisson00}, for any
% magnitude ratio as $(J_{b}/J_{a})\neq1$.

% We also study the geometric fluctuations effects on the spin 1 Heisenberg
% model when the coupling are distributed by the 6-3 sequence. This
% sequence is associated with relevant geometric fluctuations on the
% spin 1/2 chain \cite{vieira05a,vieira05b}. The substitution
% rule is given by 
% \begin{eqnarray}
% \mathbf{\sigma}_{63}:\left\{ \begin{array}{ccl}
% a & \rightarrow & babaaa\\
% b & \rightarrow & baa
% \end{array}\right.,\label{subs-rule-63}
% \end{eqnarray}
% and it leads to a sequence like $baababaaababaaabaa\dots$. the wandering
% exponent \cite{hermisson00} is
% \begin{equation}
% \omega=\frac{\ln2}{\ln5}.
% \end{equation}

The remaining of this paper is as follows. For the sake of completeness, in Sec. \ref{spin1/2-rg} 
we review the SDRG scheme of Ma, Dasgupta and Hu for the random-bond spin-$1\over 2$ Heisenberg chain.
The adaptation of the scheme to spin-$1$ chains, along the three approaches mentioned above,
is described in Sec. \ref{spin1-rg}. In Secs. \ref{fibonacci} and \ref{63-sequence} we apply
the three approaches to the Heisenberg spin-1 chain with couplings modulated by the
Fibonacci and the 6-3 sequences, which respectively induce geometric fluctuations characterized by
$\omega=0$ and $\omega\simeq 0.43$. This allows us to investigate cases representative
of different regimes of dynamic scaling in the corresponding spin-$1\over2$ chain, which
the spin-$1$ chain may be expected to approach in the strong-modulation limit.
Results are checked against QMC and DMRG simulations. Section \ref{Conc} summarizes our 
findings,
while several technical points are discussed in the appendices.

\section{\label{spin1/2-rg}Strong-disorder renormalization group for the
Heisenberg spin-$1\over 2$ chain}

The SDRG scheme of Ma, Dasgupta and Hu consists in the
iterative decimation of the strongest energy parameter --- usually
a bond connecting two spins --- and its replacement by an effective parameter
calculated by perturbation theory, in order to eliminate the highest
energy degrees of freedom present in the system. The new effective bond
is always smaller than the decimated one. After the decimation, the process
is repeated with the next strongest bond in the chain, and so on. In the asymptotic
limit of a very large number of iterations, the effective Hamiltonian generated
by this method should describe well the low-energy (low-temperature)
thermodynamic behavior of the system.

\begin{figure}
\includegraphics[width=0.7\columnwidth]{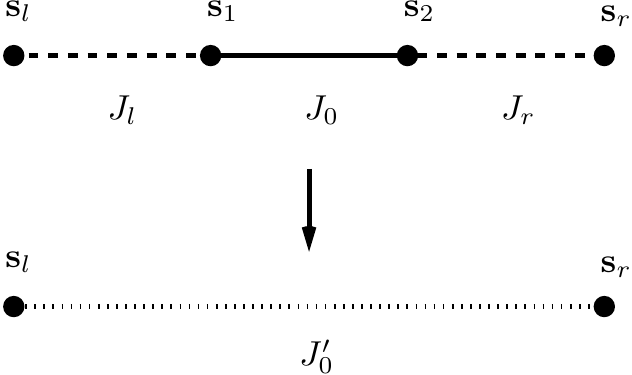} \caption{\label{rule2}Decimation procedure for a pair of 
$S={1\over 2}$ spins.}
\end{figure}

The method was first introduced to study the random-bond spin-$1\over 2$ 
Heisenberg chain,\cite{ma79,dasgupta80} and successfully reveals the thermodynamics 
properties of the corresponding ground state, which has been dubbed a random-singlet phase.
\cite{fisher94} The first step of the method is finding the strongest bond in the chain, 
say $J_{0}$. Assuming that the coupling distribution is sufficiently broad, at low 
temperatures ($T\ll J_{0}$ in suitable units), the spin pair coupled by $J_{0}$ can be pictured
as frozen in its local ground state (a singlet), and thus can be eliminated, its virtual excitations
giving rise to an effective bond coupling the spins neighboring the pair, as illustrated in 
Fig. \ref{rule2}. 

If we assume that the neighboring bonds $J_l$ and $J_r$ are much weaker that $J_{0}$,
we can calculate the effective bond by perturbation theory. Treating
the interactions between the pair and the rest of the chain, via the neighboring spins 
$\mathbf{s}_{l}$ and $\mathbf{s}_{r}$, as a perturbation
over the exact states of the pair, we can write the local Hamiltonian
as 
\begin{equation}
h = h_{0}+h_{1},\nonumber
\end{equation}
with
\begin{eqnarray}
h_{0} & = & J_{0}\mathbf{s}_{1}\cdot\mathbf{s}_{2},\nonumber \\
h_{1} & = & J_{l}\mathbf{s}_{l}\cdot\mathbf{s}_{1}+J_{r}\mathbf{s}_{2}\cdot\mathbf{s}_{r}, \nonumber
\end{eqnarray}
where $h_{1}$ represents the perturbation over the states of the pair $\mathbf{s}_{1}$
and $\mathbf{s}_{2}$ associated with $h_0$. (Throughout this paper, spin-$\frac{1}{2}$ operators
are represented in lowercase, unless explicitly stated otherwise.)
The eigenstates of $h_0$ are a singlet,
\begin{equation}
\left|\Phi_{0}\right\rangle = \frac{1}{\sqrt{2}}\left(\left|+-\right\rangle-\left|-+\right\rangle\right), 
\label{singleto}
\end{equation}
with energy $E_{0}=-\frac{3}{4}J_0$, and a triplet,
\begin{equation}\label{tripleto}
\begin{split}
\left|\Phi_{1}^{+}\right\rangle &= \left|++\right\rangle, \\
\left|\Phi_{1}^{0}\right\rangle &= \frac{1}{\sqrt{2}}\left(\left|+-\right\rangle+\left|-+\right\rangle\right), \\
\left|\Phi_{1}^{-}\right\rangle &= \left|--\right\rangle, \\
\end{split}
\end{equation}
with energy $E_{1}=\frac{1}{4}J_0$. If we assume that $h_0$ sets the 
energy scale $\Delta$ of the system, a reasonable estimate for this is 
$\Delta=E_{1}-E_{0}$, and at lower energy scales the pair $\mathbf{s}_{1}$
and $\mathbf{s}_{2}$ is effectively frozen in its ground state.

Up to second order in perturbation theory, the effective Hamiltonian is then written as
\begin{eqnarray}
h^{\mathrm{eff}} & = & \left\langle\Phi_{0}\left|h_{1}\right|\Phi_{0}\right\rangle
+\sum_{i}\frac{\displaystyle \left|\left\langle\Phi_{0}\left|h_{1}\right|\Phi_{1}^{i}\right\rangle\right|^{2}}
{\displaystyle E_{0}-E_{1}}=\nonumber \\
 & = & E^{\prime}+J_{0}^{\prime}\mathbf{s}_{l}\cdot\mathbf{s}_{r},\label{perturbation_eq}
\end{eqnarray}
with the summation running over $i\in\left\{+,0,-\right\}$.
The effective parameters are given by
\begin{equation}
E^{\prime} = -\frac{3}{4}J_{0}-\frac{3}{16}\frac{(J_{l}^{2}+J_{r}^{2})}{J_0}\quad
\mbox{and}\quad
J_{0}^{\prime} = \frac{1}{2}\frac{J_{l}J_{r}}{J_{0}}\label{eqrule2},
\end{equation}
in which $E^{\prime}$ represents a correction to the ground-state energy of $h$,
and $J_{0}^{\prime}$ is an effective coupling between the spins $\mathbf{s}_{l}$
and $\mathbf{s}_{r}$.

Notice that the effective bond $J_{0}^{\prime}$ is always smaller 
than the original bond $J_{0}$, so that the energy scale is consistently reduced.
The iteration of the above renormalization rule will lead to a probability
distribution of effective bonds, \cite{fisher94} which gets broader and broader,
suggesting that the results thus obtained are asymptotically exact.
In fact, the fixed-point probability distribution of the effective couplings
has infinite variance --- an infinite-randomness fixed point.

\section{\label{spin1-rg}Strong-disorder renormalization group for the 
Heisenberg spin-$1$ chain}

In this section, we review and discuss three different approaches to adapting
the SDRG method for spin-$1$ chains.\cite{hyman96phd,monthus97,monthus98}
The different approaches
arise from the difference between the spectrum of the spin-$1$ and 
spin-$1\over 2$ pairs, and from the search for a decimation procedure
which consistently reduces the energy scale. Other approaches have
also been considered in the literature.\cite{damle02,saguia02,igloi05} 

\subsection{The first approach}

This approach is the direct adaptation of the calculations in
the previous section to the spin-$1$ case. The local Hamiltonian is
defined by 
\begin{equation}
h=h_{0}+h_{1}, 
\end{equation}
with
\begin{eqnarray}
h_{0} & = & J_{0}\mathbf{S}_{1}\cdot\mathbf{S}_{2},\nonumber \\
h_{1} & = & J_{l}\mathbf{S}_{l}\cdot\mathbf{S}_{1}+J_{r}\mathbf{S}_{2}\cdot\mathbf{S}_{r},
\label{hamilt-local-ap0} 
\end{eqnarray}
where $h_1$ is to be treated as a perturbation over $h_0$. (Throughout this paper, spin-$1$ operators
are represented in uppercase.) The energy levels of $h_0$ are a singlet,
with energy $E_{0}=-2J_{0}$, a triplet, with energy $E_{1}=-J_{0}$, and a quintuplet, with energy
$E_{2}=J_{0}$. Discarding all excited states sets the local energy scale to $\Delta=E_{1}-E_{0}=J_{0}$.

Applying second-order perturbation theory to the above Hamiltonian, by
summing over all excited states of $h_0$, as
in Eq. \eqref{perturbation_eq}, the effective bond between $\mathbf{S}_{l}$
and $\mathbf{S}_{r}$ is given by the rule
\begin{equation}
J_{0}^{\prime}=\frac{4}{3}\frac{J_{l}J_{r}}{J_{0}},\label{effective-bond-ap0}
\end{equation}
which is not necessarily consistent, because the conditions $J_{l}<J_{0}$
and $J_{r}<J_{0}$ are not enough to guarantee that $J_{0}^{\prime}<J_{0}$.
However, this result should be valid if the coupling distribution is
sufficiently broad, i.e., if one is sure that $J_{l},J_{r}\ll J_{0}$.

As discussed below, the search for a decimation 
rule which is consistent when the above rule fails 
gives rise to two other approaches, in which the spin-$1$ pair is replaced 
by a spin-$\frac{1}{2} $ pair. 
% To complete
% the description we need to calculate the gap from the ground state
% to the first state to be decimated for each possible different block,
% this is shown in the table \ref{table-del-ap0} in appendix \ref{block-del}.
% It is possible to formulate a RG decimation proceeding for each of
% those blocks, but here for simplicity we always will avoid those decimations.

\subsection{The second approach}
\label{sec:secapp}

The second approach we discuss was used by Monthus \emph{et al.}\cite{monthus97,monthus98}
in the study of random spin-$1$ chains. The idea is to discard only the quintuplet states 
of $h_0$, by replacing the spin-$1$ pair $\mathbf{S}_{1}$
and $\mathbf{S}_{2}$ by a pair of spin-$1\over 2$ effective spins $\mathbf{s}_{1}^{\prime}$
and $\mathbf{s}_{2}^{\prime}$, also connected by a bond $J_{0}$, in order to reproduce
the lowest energy gap of $h_0$. The effective local Hamiltonian is then written as
\begin{equation}
h_{0}^{\mathrm{eff}}=-\frac{5}{4}J_{0}+J_{0}\mathbf{s}_{1}^{\prime}\cdot\mathbf{s}_{2}^{\prime}.
\end{equation}
It should be noted that the constant $-5J_{0}/4$ is used to match the states
of $h_{0}^{\mathrm{eff}}$ and $h_{0}$ in Eq. \eqref{hamilt-local-ap0}.

\begin{figure}
\includegraphics[width=0.99\columnwidth]{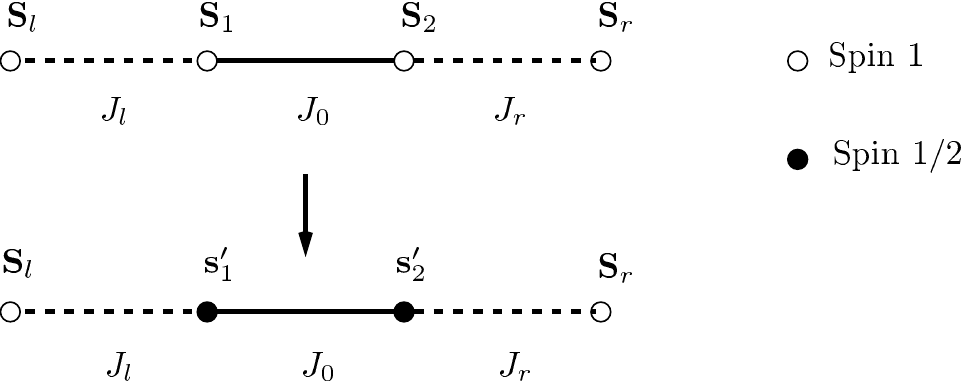} \caption{\label{rule4}
Decimation procedure for a pair of $S=1$ spins according
to the second approach.}
\end{figure}
The local Hamiltonian is replaced by an effective local
Hamiltonian with the spins $\mathbf{S}_{l}$, $\mathbf{s}_{1}^{\prime}$, 
$\mathbf{s}_{2}^{\prime}$, and $\mathbf{S}_{r}$, 
\begin{equation}
h^{\mathrm{eff}}=h_{0}^{\mathrm{eff}}+h_{1}^{\mathrm{eff}},
\end{equation}
as shown in Fig. \ref{rule4}. Now the question is how to determine the 
perturbation term $h_{1}^{\mathrm{eff}}$, which represents the connection 
between the effective spin-$\frac{1}{2}$ pair and the rest of
the chain. If one requires, to first-order in perturbation theory,
that both $h_{1}^{\mathrm{eff}}$ and $h_1$ in 
Eq. \eqref{hamilt-local-ap0} yield the same matrix elements inside
their respective singlet subspaces and inside their respective triplet subspaces,
one concludes that\cite{monthus97,monthus98}
\begin{equation}
h_{1}^{\mathrm{eff}}=J_{l}\mathbf{S}_{l}\cdot\mathbf{s}_{1}^{\prime}+
J_{r}\mathbf{s}_{2}^{\prime}\cdot\mathbf{S}_{r}\label{h-monthus}.
\end{equation}
This rule reduces the local energy scale from $3J_0$ (the gap between
the singlet and the quintuplet states of $h_{0}$) to $J_0$ (the gap
between the singlet and triplet states of $h_0^\mathrm{eff}$).

However, the effective Hamiltonian in Eq. \eqref{h-monthus}
does not reproduce the matrix elements of $h_1$ between
states in the singlet and triplet subspaces. In order to achieve this, one has
to introduce next-nearest-neighbor couplings, giving rise to the exact
first-order effective Hamiltonian\cite{hyman96phd,monthus97,monthus98}
\begin{equation}
h_{1}^{\mathrm{exact}}=J_{l}\mathbf{S}_{l}\cdot\left(\alpha_{+}
\mathbf{s}_{1}^{\prime}+\alpha_{-}\mathbf{s}_{2}^{\prime}\right)
+J_{r}\left(\alpha_{-}\mathbf{s}_{1}^{\prime}+\alpha_{+}
\mathbf{s}_{2}^{\prime}\right)\cdot\mathbf{S}_{r},\label{h-ex}
\end{equation}
with
\begin{equation}
\alpha_{\pm}=\frac{1\pm\alpha}{2}\hspace{0.5cm}\textrm{and}\hspace{0.5cm}\alpha=
\sqrt{\frac{8}{3}}\simeq 1.63.
\end{equation}

As the next-nearest-neighbor bonds are ferromagnetic,
$\alpha_{-}J_{l,r}\simeq -0.316J_{l,r}$, they do not introduce frustration 
in the system. 
Note also that, although the nearest-neighbor effective bonds, 
$\alpha_{+}J_{l,r}\simeq1.32J_{l,r}$, are stronger than the original ones,
it can be checked that
the associated gap of the four-spin cluster decreases as compared
to $3J_0$ (see Tab. \ref{table-del-ap1} in App. \ref{block-del}), 
so that the energy scale is still consistently reduced.

Due to the nonfrustrating character of the ferromagnetic bonds, Monthus \emph{et al.}
argued that it is safe to ignore them, if one is interested only in qualitative features
of the physical effects introduced by randomness, and build a renormalization scheme
based on the effective Hamiltonian of Eq. \eqref{h-monthus}. This forms the basis
for the second approach.
% The equation \eqref{h-monthus}, $h_{1}^{\mathrm{eff}}$, gives rise to what
% we call from now on the approach 1. The complete perturbative term,
% $h_{1}^{\mathrm{exact}}$, equation \eqref{h-ex}, gives rise to the approach
% 2. We discuss now the approach 1, and the approach 2 will be discussed
% in the next subsection.

\begin{figure}
\includegraphics[width=0.7\columnwidth]{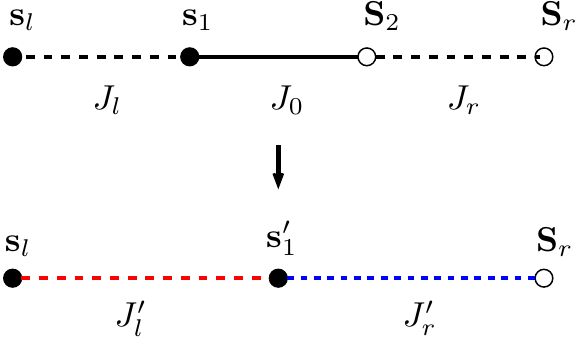} 
\caption{\label{rule3}Decimation procedure for a mixed-spin pair according
to the second approach.}
\end{figure}
Since the effective perturbative term introduces spin-$\frac{1}{2}$ objects,
one needs to deal with renormalization steps involving both spin-$1$ and
spin-$1\over2$ operators in order to have a closed scheme for the renormalization
group. 
% Although we are concerned here with the case of strong modulation,
% in which the situations described below will in practice not appear, we address
% them for the sake of completeness.

There is clearly the possibility that the largest local energy scale is 
set by a pair composed of
a spin-$1\over 2$ object $\mathbf{s}_1$ and a spin-$1$ object $\mathbf{S}_2$ 
connected by a bond $J_0$, and interacting with neighboring spins 
$\mathbf{s}_l$ and $\mathbf{S}_r$ via weaker bonds $J_l$ and $J_r$, as shown
in Fig. \ref{rule3}.
The ground state of the pair corresponds to a doublet, giving rise to 
an effective spin-$1\over2$ object $\mathbf{s}_1^{\prime}$, and to first order in 
perturbation theory the four-spin cluster can be described by an effective Hamiltonian
\begin{equation}
h^{\mathrm{eff}}=J_{l}^{\prime}\mathbf{s}_{l}\cdot\mathbf{s}_{1}^{\prime}+
J_{r}^{\prime}\mathbf{s}_{1}^{\prime}\cdot\mathbf{S}_{r},
\label{eqrule3-1}
\end{equation}
with
\begin{equation}
J_{l}^{\prime} = -\frac{1}{3}J_{l}\qquad\mbox{and}\qquad
J_{r}^{\prime} = \frac{4}{3}J_{r}.
\label{eqrule3-2}
\end{equation}

\begin{figure}
\includegraphics[width=0.7\columnwidth]{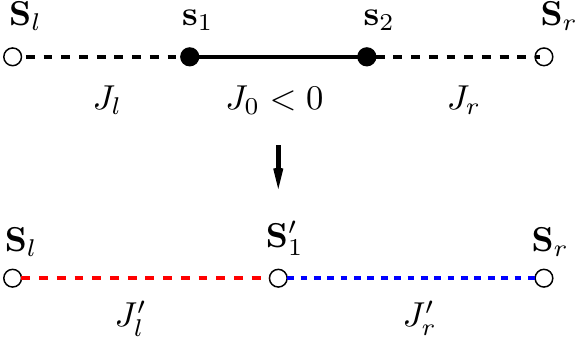} \caption{\label{rule1}
Decimation procedure for a pair of ferromagnetically coupled spin-$1\over2$ objects according
to the second approach.}
\end{figure}
Notice that this last process generates ferromagnetic bonds, but these only connect
spin-$1\over2$ objects. In case the local energy scale is set by such a bond, $-\left|J_0\right|$,
connecting $\mathbf{s}_1$ and $\mathbf{s}_2$, the unperturbed ground state is
a triplet, giving rise to an effective spin-$1$ object $\mathbf{S}_1^{\prime}$; see Fig. \ref{rule1}. 
A first-order perturbative calculation leads to an effective Hamiltonian
\begin{equation}
h^{\mathrm{eff}}=J_{l}^{\prime}\mathbf{S}_{l}\cdot\mathbf{S}_{1}^{\prime}+
J_{r}^{\prime}\mathbf{S}_{1}^{\prime}\cdot\mathbf{S}_{r},
\quad\mbox{with}\quad J_{l,r}^{\prime}=\frac{1}{2}J_{l,r}.
\label{eqrule1}
\end{equation}

% For the both previous cases, it was not needed to discuss any kind
% of different approach to decimate the largest parameter because there
% are only two energy levels for the decimated pair.

To summarize, in this second approach there are four kinds of bonds, and each
of them requires a different decimation rule. In the same notation
used in Ref. \onlinecite{monthus98}, these are:
%\begin{itemize}

\textit{(i) Rule 1}: A pair of $S={1\over2}$ spins connected by an ferromagnetic
bond [Fig. \ref{rule1}, Eq. (\ref{eqrule1})]; 

\textit{(ii) Rule 2}: A pair of $S={1\over2}$ spins connected by an antiferromagnetic
bond [Fig. \ref{rule2}, Eqs. (\ref{perturbation_eq}) and (\ref{eqrule2})];

\textit{(iii) Rule 3}: A mixed-spin pair connected by
an antiferromagnetic bond [Fig. \ref{rule3}, Eqs. (\ref{eqrule3-1}) and (\ref{eqrule3-2})];

\textit{(iv) Rule 4}: A pair of $S=1$ spins connected by an antiferromagnetic
bond [Fig. \ref{rule4}, Eq. (\ref{h-monthus})]. 
%\end{itemize}

Which rule is to be applied depends on which bond sets the energy
scale at a given step of the SDRG scheme. Using as an estimate for such a scale
the local gap $\Delta$ between the ground state and the first discarded excited 
energy level of the spin pair, we have for the different rules
\begin{eqnarray}
\Delta_{1} & = & -J_{0}=|J_{0}|,\nonumber \\
\Delta_{2} & = & J_{0},\nonumber \\
\Delta_{3} & = & \tfrac{3}{2} J_{0},\nonumber \\
\Delta_{4} & = & 3J_{0}.
\label{delta1}
\end{eqnarray}
The SDRG scheme for the random-bond spin-$1$ chain then amounts to recursively looking 
for the bond associated with the largest
$\Delta$ and applying the corresponding decimation rule.

In the case of deterministic aperiodicity generated by substitution rules, 
there appear blocks composed of more than one strong bond, as in the Fibonacci
sequence ($abaababaabaab\dots$) with $J_a>J_b$. In order to deal with these cases, the set of decimation
rules has to be extended, as described, in the spin-$1\over2$ case, in Refs.
\onlinecite{vieira05a,vieira05b}. The starting point is to find the spin block
yielding the largest local energy gap, and renormalizing it to either an effective
bond between the spins neighboring the block, or to one or two effective spins,
according to the lowest energy levels of the original block; see App. \ref{block-del}. 
The effective couplings are then to be calculated by first- or second-order perturbation theory.
In order to avoid such complications as much as possible, we choose $J_b>J_a$ and focus on the
strong-modulation regime for analytical calculations. Moderate modulation can be studied
numerically using SDRG, by implementing the rules for different blocks, and results from such calculations
are briefly mentioned below.

% Even though we need to know the gap $\Delta$ to first excited state
% to compare with the $J_{b}$ single block gaps. We performed this
% calculation numerically for blocks with different sizes and spin configurations,
% and the results are presented in the appendix \ref{block-del}.

\subsection{The third approach}
\label{sec:thirdapp}

\begin{figure}
\includegraphics[width=0.99\columnwidth]{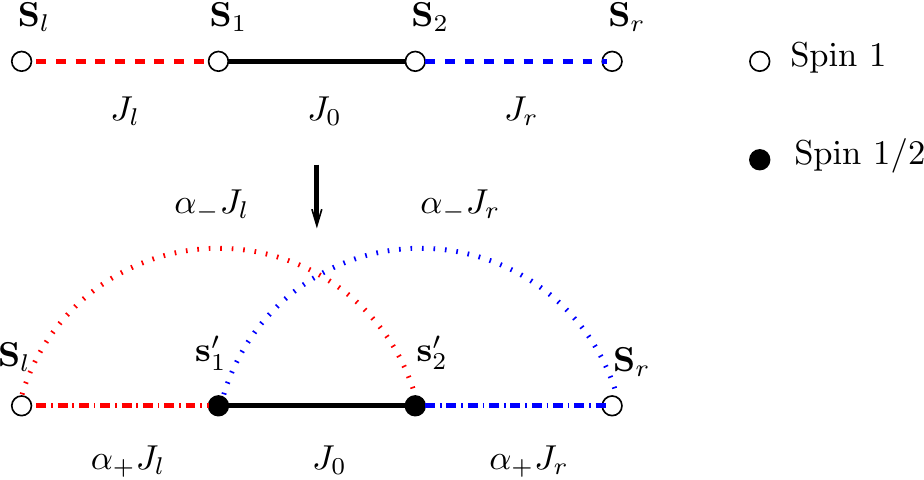} \caption{\label{rule4prime}
Decimation procedure for a pair of $S=1$ spins according
to the third approach.}
\end{figure}
The third approach consists in using the exact first-order Hamiltonian
$h_1^{\mathrm{exact}}$ of Eq. \eqref{h-ex}  as the effective local Hamiltonian
that arises from the decimation of a spin-$1$ pair. Among the decimation
rules, of concern here is a modification of rule 4 along the lines of 
Eq. (\ref{h-ex}). With the introduction of both nearest- and next-nearest-neighbor
bonds, it is possible that a spin $\mathbf{S}_1$ is strongly coupled to a spin 
$\mathbf{S}_2$ while both are weakly coupled to a number of other spins. Therefore, the exact
first-order Hamiltonian turns into
\begin{eqnarray}
h_{1}^{\mathrm{exact}} & = & \sum_{i=1}^{n_l} J_{l}^{(i)}\mathbf{S}_{l}^{(i)}\cdot\left(\alpha_{+}
\mathbf{s}_{1}^{\prime}+\alpha_{-}\mathbf{s}_{2}^{\prime}\right) \nonumber \\
& + & \sum_{i=1}^{n_r}J_{r}^{(i)}\left(\alpha_{-}\mathbf{s}_{1}^{\prime}+\alpha_{+}
\mathbf{s}_{2}^{\prime}\right)\cdot\mathbf{S}_{r}^{(i)} \label{h-ex-m},
\end{eqnarray}
where $n_l$ is the number of spins $\mathbf{S}_l^{(i)}$ to which
$\mathbf{S}_{1}$ is weakly coupled via $J_{l}^{(i)}$, and 
$n_r$ is the number of spins $\mathbf{S}_r^{(i)}$ to which
$\mathbf{S}_{2}$ is weakly coupled via $J_{r}^{(i)}$.
Thus, rule 4 now reads
%\begin{itemize}

{\textit{(iv') Rule} $4^\prime$: A pair of $S=1$ spins connected by an antiferromagnetic
bond [Fig. \ref{rule4prime}, Eq. \eqref{h-ex-m}].} 
%\end{itemize}
Notice that Fig. \ref{rule4prime} illustrates the case $n_l=n_r=1$.

\section{\label{fibonacci}The spin-$1$ Fibonacci-Heisenberg chain}

The Fibonacci sequence is produced by the iterative application of the
substitution rule
\begin{equation}
\sigma_{\textrm{fb}}:\left\{ \begin{array}{l}
a\to ab\\
b\to a\end{array}\right.,
\label{eq:fibsubrule}
\end{equation}
starting from a single letter (either $a$ or $b$).

The spin-$1\over2$ Heisenberg chain with couplings $J_i\in\{J_a,J_b\}$ 
following the Fibonacci sequence remains critical (i.e. gapless) 
for all finite values of the coupling ratio $J_b/J_a$. Since
enforced dimerization makes the chain gapped, for general aperiodic
sequences the relevant geometric fluctuations to be measured are those associated
with pairs of subsequent letters. As discussed in Ref. \onlinecite{vieira05b}
(and references therein),
these grow with the chain length as a power-law, with a wandering exponent
$\omega$ related to the substitution rule for letter pairs,
rather than for single letters. (For an example concerning the Fibonacci
sequence, see App. \ref{pairomega} )

In the case of the Fibonacci sequence, this exponent is $\omega=0$,
in contrast to the random-bond chain, for which $\omega={1\over2}$.
Thus, geometric fluctuations associated with couplings chosen from the
Fibonacci sequence are much weaker than those produced by a random
coupling distribution. Despite this fact, Fibonacci couplings also induce
dramatic changes in the low-temperature behavior of the Heisenberg
spin-$1\over 2$ chain.\cite{vieira05a,vieira05b}
As we show below, this is not the case for the Heisenberg spin-$1$ chain.

In the following sections, we present the results of applying the 
three different SDRG approaches defined in the previous section to the 
problem of the Fibonacci-Heisenberg spin-$1$ chain. We also discuss
the discrepancies between the second and the other two approaches,
and present results from quantum Monte Carlo and DMRG calculations,
which point to the fact that, in contrast to the random-bond chain,
the second approach does lead to qualitatively incorrect conclusions
about the low-temperature behavior of the system.

\subsection{SDRG: The first approach}

\begin{figure}
\includegraphics[width=\columnwidth]{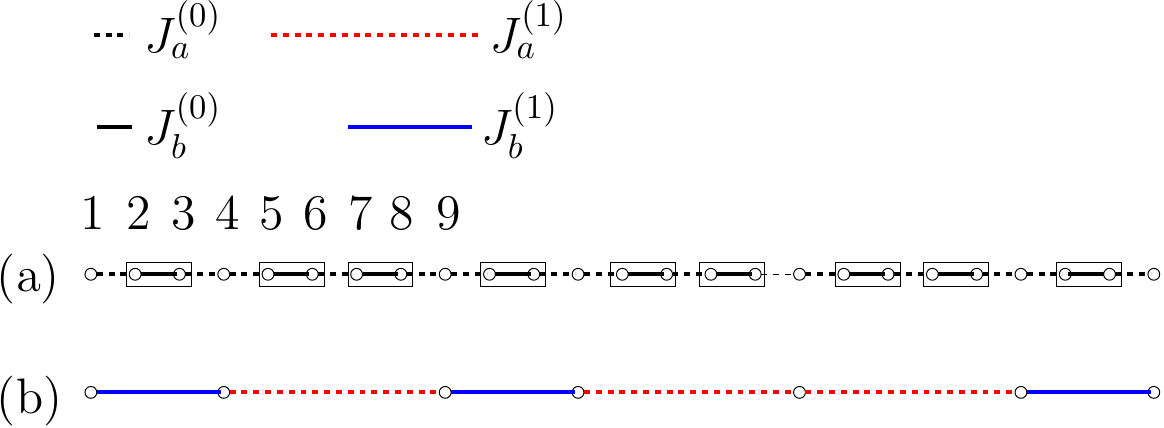} 
\caption{\label{fibonacci-chain-ap0}
(a) Coupling distribution of the
spin-$1$ Fibonacci-Heisenberg chain.
(b) Effective chain obtained from the first SDRG approach
after a single lattice sweep.}
\end{figure}

Figure \ref{fibonacci-chain-ap0}(a) shows the first few bonds near the
left end of the Fibonacci-Heisenberg chain. As mentioned before,
throughout the paper we assume $J_{b}>J_{a}$, but here, in order
to apply the first approach, we assume the stronger condition
$J_b\gg J_a$.

According to the usual recipe of the first approach,
all $J_{b}$ bonds, which appear enclosed in Fig. \ref{fibonacci-chain-ap0}(a),
are to be decimated in a first SDRG lattice sweep, giving rise to effective couplings.
Between spins 1 and 4 in Fig. \ref{fibonacci-chain-ap0}(a) 
there is one spin pair connected by an isolated $J_{b}$ bond,
and its decimation results in an effective bond $J_{b}^{\prime}$,
by directly applying Eq. (\ref{effective-bond-ap0}).
But there is also another effective bond, $J_{a}^{\prime}$,
which appears for instance between spins 4 and 9
by sequentially decimating the $J_b$ bonds connecting
spins $5$-$6$ and $7$-$8$. Thus we have
\begin{eqnarray}
J_{a}^{\prime}=\left(\frac{4}{3}\right)^{2}\frac{J_{a}^{3}}{J_{b}^{2}}
\quad\mbox{and}\quad
J_{b}^{\prime}=\frac{4}{3}\frac{J_{a}^{2}}{J_{b}}.
\label{eq:fibrec1stap}
\end{eqnarray}
Notice that if the effective bond $J_b^{\prime}$ is to be smaller than 
the original bond $J_b$, so that the decimations lead to a reduction of the energy scale, 
we must have $J_{b}>\sqrt{4/3}J_{a}$, which constitutes a consistency condition
for the first approach.

% We point out that the sequentially application of the equation \eqref{effective-bond-ap0}
% is only possible because the first decimation causes no changes in
% the second $J_{b}$ bond, we will discuss more about this issue in
% the appendix \ref{third-order-pt}.

As hinted by Fig. \ref{fibonacci-chain-ap0}(b), decimating all original
$J_b$ bonds leads to a Fibonacci modulation of the effective bonds
(disregarding the first effective bond as a boundary effect). It is then 
clear that a new SDRG sweep will again generate a Fibonacci sequence,
and so on. Therefore we can define
recursive equations for the effective parameters, as well as for the ratio between 
them. These are given by
\begin{eqnarray}
J_{a}^{(j+1)} & = & \left(\frac{4}{3}\right)^{2}\frac{\left[J_{a}^{(j)}\right]^{3}}{\left[J_{b}^{(j)}\right]^{2}},
\quad J_{b}^{(j+1)} = \frac{4}{3}\frac{\left[J_{a}^{(j)}\right]^{2}}{J_{b}^{(j)}},
\nonumber \\
r^{(j+1)} & \equiv & \frac{J_{b}^{(j+1)}}{J_{a}^{(j+1)}}=\left(\frac{3}{4}\right)r^{(j)},
\label{eq:fibap0rec}
\end{eqnarray}
in which $j$ labels the SDRG lattice sweep, $j=0$ corresponding to the original chain.

Notice that the ratio between the effective bonds decreases along the RG 
iterations. This means that the effective chain looks more and more 
uniform as the energy scale is reduced, and we then conclude that a 
Fibonacci modulation does not drive 
the system towards an aperiodic singlet phase even for an arbitrarily large
initial coupling ratio, unlike what is observed for the Fibonacci-Heisenberg
spin-$1\over2$ chain.\cite{vieira05a,vieira05b}

Thus we expect the chain to remain in the Haldane phase, but with a gap
which depends on the bare coupling ratio $r=J_b/J_a$. An estimate of this gap
is provided by the value of the effective couplings at the energy scale
for which the effective coupling ratio becomes of order $1$. This 
happens after $j^*$ iterations of the SDRG scheme, with
\begin{equation}
j^* = \frac{\ln r}{\ln\frac{4}{3}}.
\end{equation}
From the
above equations, we thus conclude that the gap should behave as
\begin{equation}
 \Delta\left(r\right)\sim r^{-\frac{\ln r}{\ln(4/3)}} J_b,
 \label{eq:fibgap}
\end{equation}
up to a multiplicative constant.
Taking logarithms on both sides, this last result can be rewritten as
\begin{equation}
\ln\Delta\left(r\right)\sim \ln J_b - \frac{\ln^2 r}{\ln(4/3)}, 
\end{equation}
making evident that the gap vanishes asymptotically as the bare coupling ratio 
becomes larger and larger, with $J_b$ held constant.

It is also possible to follow the growth of bond lengths as the SDRG scheme
proceeds.
If we denote by $\ell_a^{(j)}$ and $\ell_b^{(j)}$ the respective lengths 
of the weak and strong bonds after $j$ SDRG lattice sweeps, inspection of 
Fig. \ref{fibonacci-chain-ap0} leads to relations which can be written
in matrix form as
\begin{eqnarray}
\left[\begin{array}{c}
\ell_{a}^{(j+1)}\\
\ell_{b}^{(j+1)}\\
\end{array}\right]=\left[\begin{array}{cc}
3 & 2 \\
2 & 1 \\
\end{array}\right]\cdot\left[\begin{array}{c}
\ell_{a}^{(j)}\\
\ell_{b}^{(j)}\\
\end{array}\right],
\end{eqnarray}
so that the asymptotic growth of the bond lengths follows
\begin{equation}
 \ell_a^{(j)}\sim \ell_b^{(j)}\sim \tau^j, 
\end{equation}
with $\tau=2+\sqrt{5}$, the largest eigenvalue of the above matrix,
corresponding to the rescaling factor of the renormalization-group
transformation.
Taking into account the bare lengths $\ell_a^{(0)}=\ell_b^{(0)}=1$, the
asymptotic length of the strong bonds is given by
\begin{equation}
  \ell_b^{(j)}\simeq\frac{1+\sqrt{5}}{2\sqrt{5}}\tau^j\equiv c\tau^j.
\end{equation}

An estimate for the correlation length of the spin-$1$ Fibonacci-Heisenberg chain is provided
by the length of the strong bonds at the SDRG iteration where
the effective coupling ratio becomes of order 1. Thus, we have
\begin{equation}
  \xi \sim  \ell_b^{\left(j^*\right)} \simeq c r^\nu,
  \label{eq:corrlengthfib}
\end{equation}
showing that the correlation length diverges at the infinite-modulation
limit as a power law with a quite large exponent
\begin{equation}
  \nu = \frac{\ln\tau}{\ln{\frac{4}{3}}}\simeq 5.02.
  \label{eq:nufib}
\end{equation}

\subsection{SDRG: The second approach}

Now we study the 
conclusions we can extract from the second approach by applying it
to the strong-modulation case $J_b \gg J_a$.

\begin{figure}
\includegraphics[width=0.99\columnwidth]{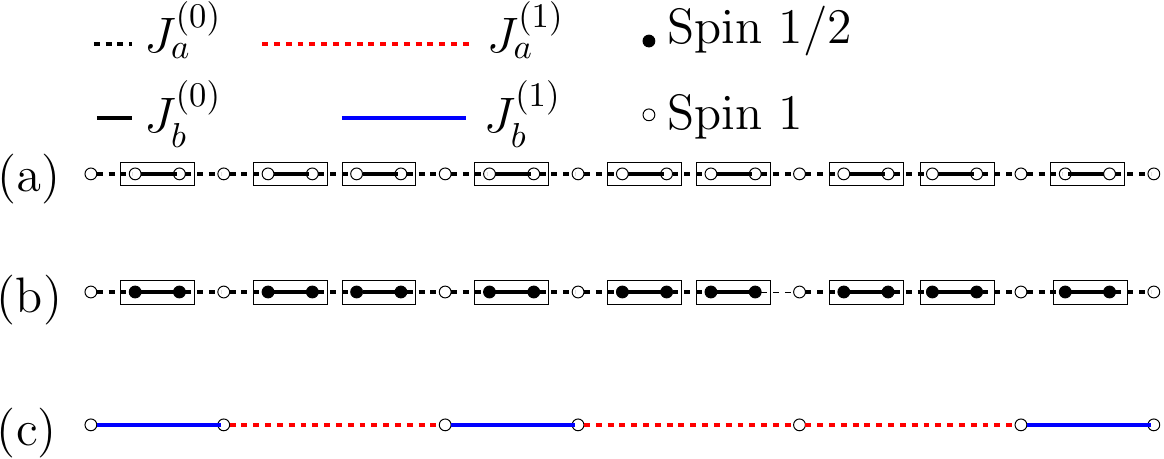} 
\caption{\label{fibonacci-chain-ap1}
Renormalization of 
the spin-$1$ Fibonacci-Heisenberg chain
according to the second SDRG approach. (a) The original chain.
(b) Spin-$1$ pairs connected by (strong) $J_b$ bonds are
replaced by spin-$1\over 2$ pairs. 
(c) Spin-$1\over2$ pairs are decimated, yielding effective couplings between
remaining $S=1$ spins.
}
\end{figure}

Figure \ref{fibonacci-chain-ap1} pictures the steps required
to obtain effective couplings in the Fibonacci-Heisenberg chain according
to the second approach. The original chain is shown in 
Fig. \ref{fibonacci-chain-ap1}(a). Applying rule 4 of Sec. \ref{sec:secapp} to all $J_b$ bonds 
connecting spin-$1$ pairs, these are replaced by spin-$1\over 2$
pairs, as shown in Fig. \ref{fibonacci-chain-ap1}(b). As we assume
$J_b\gg J_a$, the next step involves decimating all $J_b$ bonds 
connecting spin-$1\over2$ pairs, yielding effective couplings
\begin{equation}
J_{a}^{\prime}=\left(\frac{1}{2}\right)^{2}\frac{J_{a}^{3}}{J_{b}^{2}}
\quad\mbox{and}\quad J_{b}^{\prime}=\frac{1}{2}\frac{J_{a}^{2}}{J_{b}}.
\end{equation}
Again, ignoring the leftmost bond in Fig. \ref{fibonacci-chain-ap1}(c),
the effective couplings follow a Fibonacci sequence. 

From the above equations, it is clear that the values of the effective
couplings predicted by the second approach are significantly smaller
than the ones predicted by the first approach. This fact leads to errors 
when using effective $J_b$ couplings to estimate the energy levels of the 
Fibonacci-Heisenberg chain, but also, in contrast to the first approach, 
it is clear that the effective coupling ratio predicted by the second
approach,
\begin{equation}
r^{\prime}=\frac{J_{b}^{\prime}}{J_{a}^{\prime}}=2\frac{J_{b}}{J_{a}}=2r,
\end{equation}
is \emph{larger} than the bare coupling ratio $r$. Therefore, according
to the second approach, the effective coupling ratio should become
larger and larger as the SDRG scheme is iterated, so that Fibonacci-modulated 
couplings should induce
an aperiodicity-dominated gapless phase analogous to the one
observed for the Fibonacci-Heisenberg spin-$1\over 2$ chain.
This conclusion is qualitatively incorrect, since, as we will see below, 
taking into account the next-nearest-neighbor
bonds neglected in the second approach recovers the predictions of the
first approach for strong modulation.

\subsection{SDRG: The third approach}
\label{fibonacci-ap2}

\begin{figure}
\includegraphics[width=0.99\columnwidth]{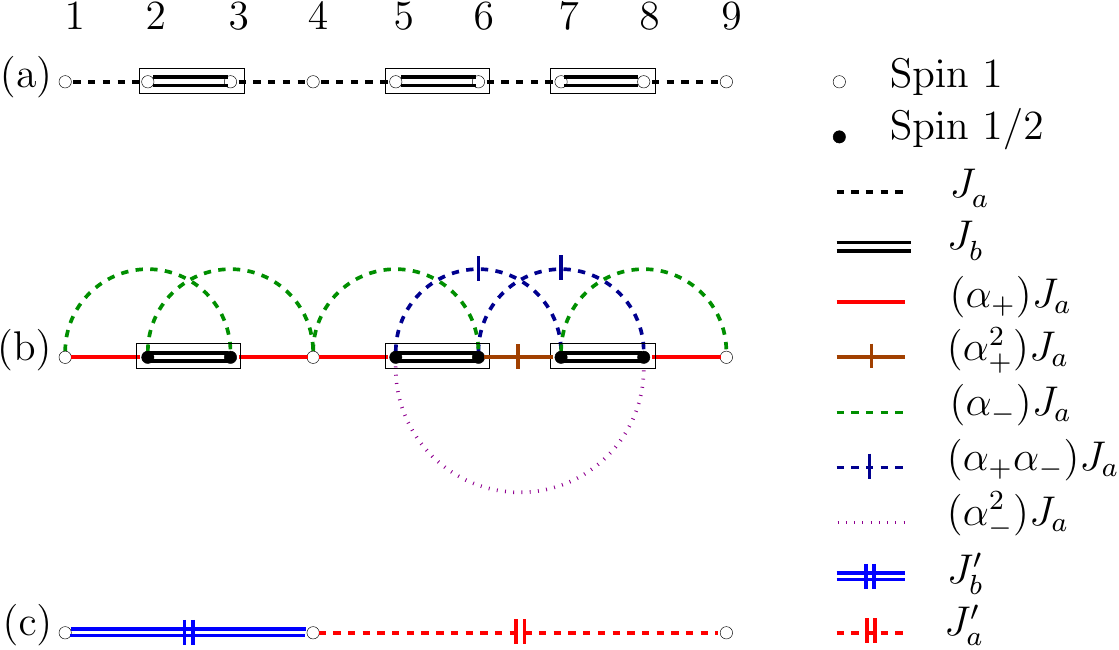} \caption{\label{fibonacci-boundary}
Renormalization of 
the spin-$1$ Fibonacci-Heisenberg chain
according to the third SDRG approach. (a) The original chain.
(b) Spin-$1$ pairs connected by (strong) $J_b$ bonds are
replaced by spin-$1\over 2$ pairs, are further-neighbor couplings are
produced. 
(c) Spin-$1\over2$ pairs are decimated, yielding effective couplings between
remaining $S=1$ spins.}
\end{figure}

When applying the third approach to the Fibonacci-Heisenberg chain following the recipe of Secs. \ref{sec:secapp} and \ref{sec:thirdapp}, after replacing all spin-$1$ pairs connected by $J_b$ bonds by 
spin-$1\over2$ pairs, there appear next-nearest- and further-neighbor bonds
as illustrated in Fig. \ref{fibonacci-boundary}(b). In particular,
the coupling between spins $5$ and $8$ in Fig. \ref{fibonacci-boundary}(b)
appears due to the repeated application of rule $4^\prime$. 

For $J_b>\alpha_+^2 J_a\simeq 1.73\,J_a$, the largest local gap in Fig. 
\ref{fibonacci-boundary}(b) is provided by the nearest-neighbor $J_b$
bonds (see App. \ref{block-del}), 
which should then be decimated to yield the effective couplings
shown in Fig. \ref{fibonacci-boundary}(c). This procedure is different
for the $J_b$ bonds which are separated from other $J_b$ bonds by at least
two weaker $J_a$ bonds (such as the bond between spins 2 and 3 in the
figure) and for the $J_b$ bonds separated by a single $J_a$ bond 
(as in the sequence of bonds between spins 5 and 8). 

In the former,
case we have to treat all weaker bonds (nearest and next-nearest) as
perturbations over the Hamiltonian
\begin{equation}
h_0=J_b \mathbf{s}_{2}\cdot\mathbf{s}_{3},
\end{equation}
following a second-order perturbative approach analogous to the one
in Eq. \eqref{perturbation_eq}. The result is an effective bond
between spins 1 and 4 in Fig. \ref{fibonacci-boundary}, given by
\begin{equation}
 J_b^\prime = \frac{\left(\alpha_+-\alpha_-\right)^2}{2}\frac{J_a^2}{J_b} = 
 \frac{4}{3}\frac{J_a^2}{J_b}.
\end{equation}

In the latter case, so that we avoid ambiguities arising from the order
in which the $J_b$ bonds are decimated, we must perform a third-order
perturbative calculation in which all weaker bonds (nearest and 
next-nearest) are treated as perturbations over the Hamiltonian
\begin{equation}
h_0=J_b \mathbf{s}_{5}\cdot\mathbf{s}_{6}+ 
J_b \mathbf{s}_{7}\cdot\mathbf{s}_{8}.
\label{h0-twopairs}
\end{equation}
As detailed in App. \ref{third-order-pt}, this yields an 
effective bond connecting spins 4 and 9, given by
\begin{equation}
 J_a^\prime = \frac{\left(\alpha_+-\alpha_-\right)^4}{4} \frac{J_a^3}{J_b^2} = 
 \left(\frac{4}{3}\right)^2\frac{J_a^3}{J_b^2}.
 \label{eq:fibjaprime}
\end{equation}

Thus, comparing the above results with Eq. \eqref{eq:fibrec1stap}, we
see that for $J_b > \alpha_+^2 J_a$ the third approach yields 
exactly the same effective bonds as the first approach. Therefore, properly taking
into account next-nearest neighbor bonds generated by the SDRG scheme
fixes the qualitatively incorrect prediction of the second approach
that strong Fibonacci modulations induce a gapless, aperiodicity-dominated phase in 
the Heisenberg spin-$1$ chain.

For weaker coupling ratios, $1<J_b/J_a<\alpha_+^2$, the largest local gap in Fig.
\ref{fibonacci-boundary}(b) is not set by the $J_b$ bonds, and the order of the
decimations is altered. Numerical implementations of the third approach indicate
that the distribution of effective bonds becomes dimerized. As it is known that 
a dimerized spin-$1\over 2$ chain has a ground state which is adiabatically connected
to the Haldane phase\cite{hida92}, this is in qualitative agreement with the
prediction of the first approach, and with the fact that the Haldane phase
is stable towards Fibonacci modulations for any value of the coupling ratio.

\subsection{\label{sec:FibQMC}Comparison with QMC simulations}

According to the SDRG predictions, for strong modulation ($r\gg 1$), 
the Fibonacci-Heisenberg 
chain can be approximated as a collection of independent spin pairs,
coupled in singlet states by the effective $J_b$ bonds. 
In the spin-$1\over2$ case, for which the ground state is expected to be
in the aperiodic singlet phase,\cite{vieira05b} this picture
should be qualitatively correct at all temperatures, provided
the modulation is strong enough. On the other hand, in the spin-$1$
case, this picture should break down at temperatures below the energy
gap $\Delta\left(r\right)$ of Eq. \eqref{eq:fibgap}. Therefore,
we can estimate the free energy of the Fibonacci-Heisenberg chain
as 
\begin{equation}
f(B,T)=\frac{1}{2}\sum\limits _{j=0}^{j^*}(n_{j}-n_{j+1})F_{\mathrm{pair}}
\left(J_{b}^{(j)};B,T\right),
\label{eq:findsing}
\end{equation}
where $F_{\mathrm{pair}}\left(J_{b}^{(j)};B,T\right)$ is the free energy of
a spin pair interacting via the Hamiltonian
\begin{equation}
H_{\mathrm{pair}}=J_{b}^{(j)}\mathbf{S}_{1}\cdot\mathbf{S}_{2}-B(S_{1}^{z}+S_{2}^{z})
\end{equation}
(with $B$ a small magnetic field, introduced to allow the calculation of the 
magnetic susceptibility), $n_j$ is the fraction of active spins 
(those not yet decimated)
at the $j$-th iteration of the SDRG scheme, and $j^*$ is the iteration at
which the effective coupling ratio becomes of order unity. From the 
above discussion, it is clear that $j^*=\infty$ for the spin-$1\over2$ case,
while it can be shown from Eqs. \eqref{eq:fibap0rec} that 
$j^*=\ln r/\ln\left(4/3\right)$ for the spin-$1$ case.

For the first $j^*$ iterations of the SDRG scheme, as the effective couplings
always follow a Fibonacci sequence, the fraction of active spins satisfy the
recurrence relation $n_{j+1}=(1-2f_{b})n_{j}$, where 
$f_{b}=(3-\sqrt{5})/2\simeq 0.382$
is the fraction of letters $b$ in the infinite Fibonacci sequence
(see App. \ref{fibonacci-appendix}). Thus, we obtain $n_j = (1-2f_b)^j$.

The susceptibility at zero field is readily obtained from the free energy,
\begin{equation}
\chi(T)=-\left.\frac{\partial^{2}f}{\partial B^{2}}\right|_{B=0}.
\label{eq:suscindsing}
\end{equation}

\begin{figure}
\includegraphics[width=0.99\columnwidth]{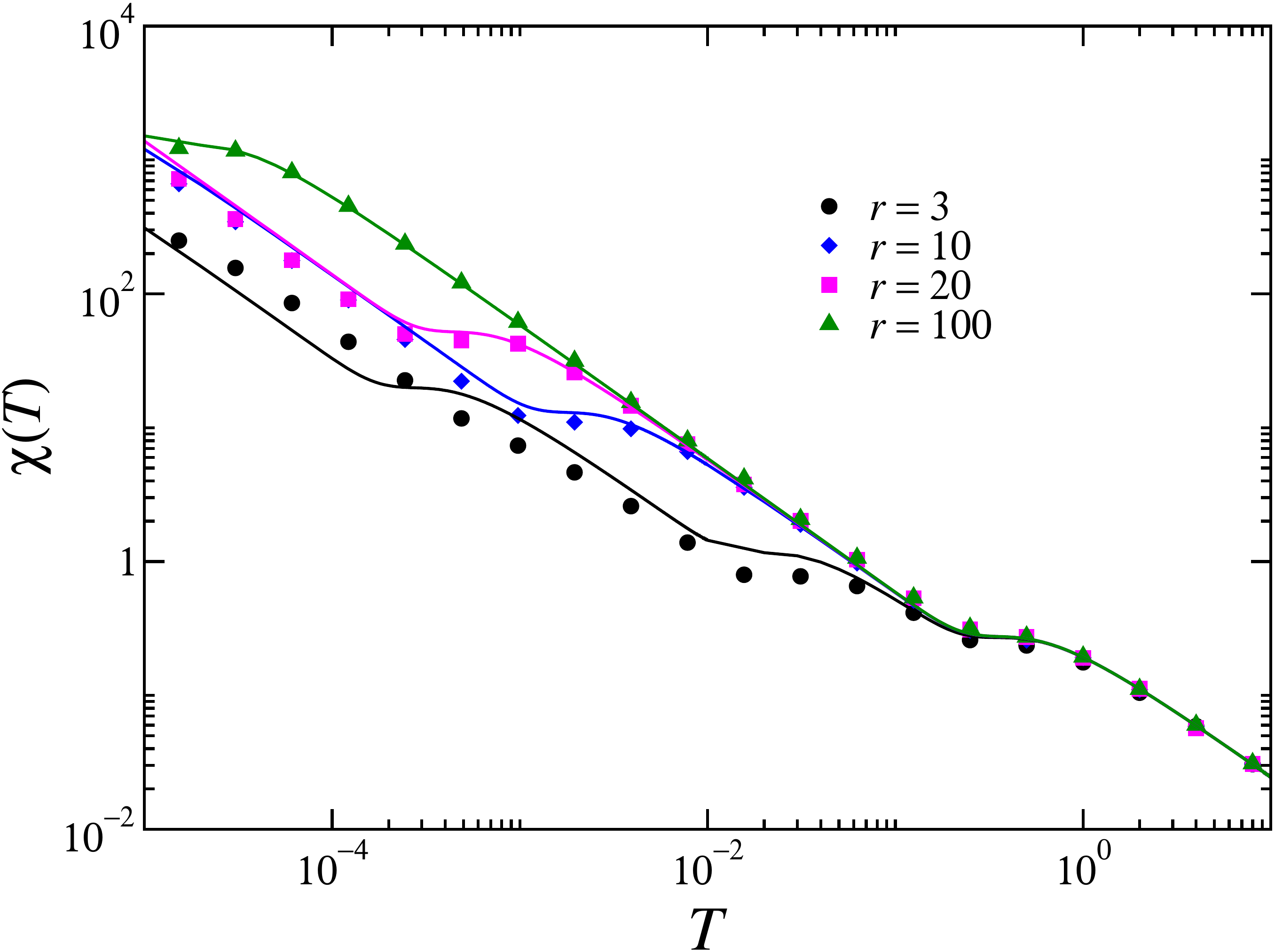}
\caption{\label{chi_of_T_Fibonacci_L90_OBC_S1.2}Magnetic susceptibility 
as a function of temperature for
the spin-1/2 Fibonacci-Heisenberg chain. Solid lines correspond to the 
SDRG prediction for different coupling ratios $r=J_b/J_a$. 
QMC results (symbols) 
were obtained using chains with 90 sites and open boundary conditions. Error bars are smaller than symbol size.}
\end{figure}
We first checked the SDRG predictions for the spin-$1\over2$ chain, 
using the effective couplings calculated in Ref. \onlinecite{vieira05b},
by comparing the results of Eqs. \eqref{eq:findsing} and \eqref{eq:suscindsing} 
with QMC simulations, performed using the stochastic series expansion scheme~\cite{Sandvik91,Sandvik92} 
with directed loop updates~\cite{Syljuasen02}. As shown in Fig. \ref{chi_of_T_Fibonacci_L90_OBC_S1.2},
the SDRG prediction gets closer and closer to the QMC results as the
modulation increases, as expected from the perturbative nature of the SDRG scheme.

\begin{figure}
\includegraphics[width=0.99\columnwidth]{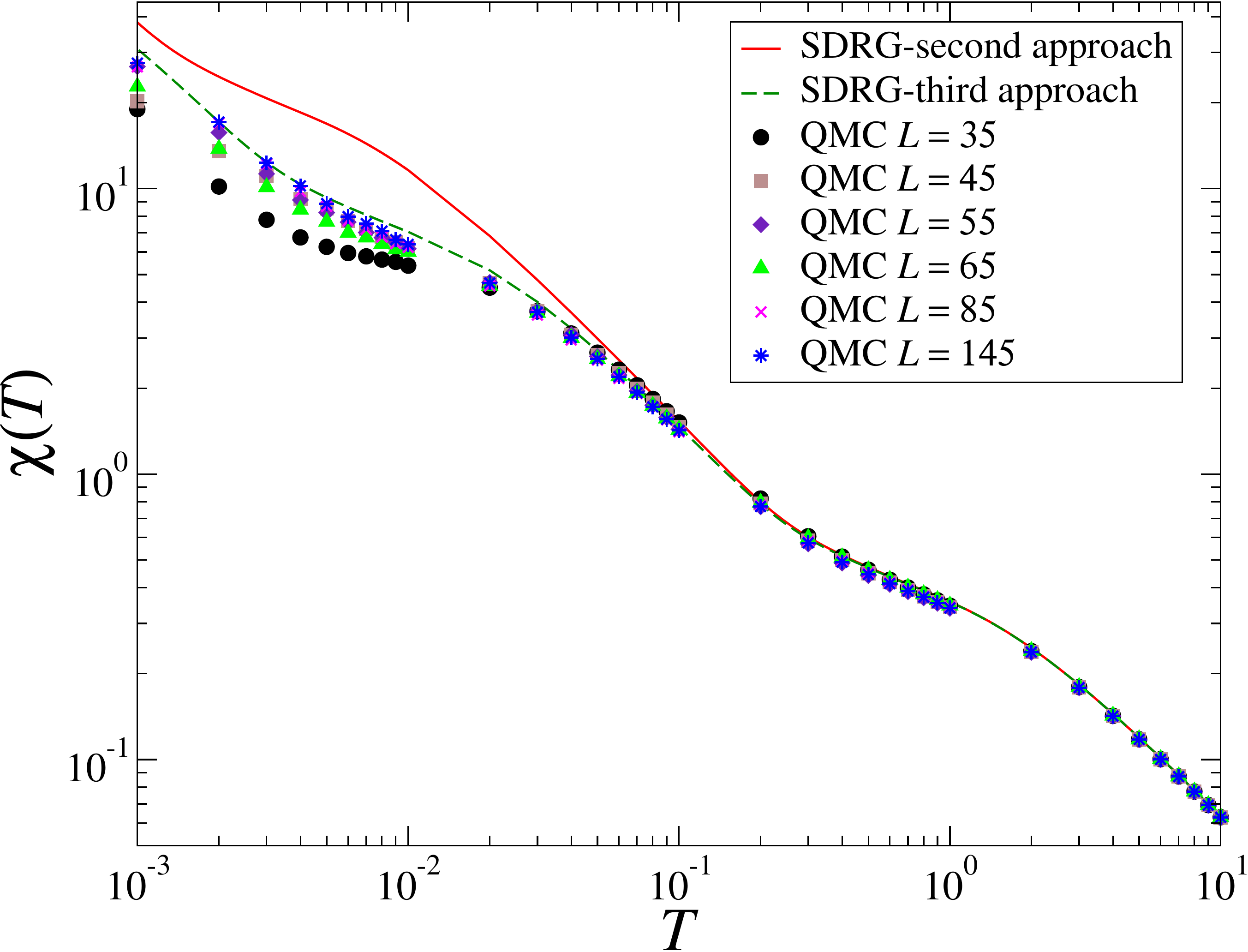} 
\caption{\label{chi-fibonacci-r10}Temperature dependence of the magnetic susceptibility
for the spin-$1$ Fibonacci-Heisenberg chain with coupling ratio $r=10$. Solid (dashed) 
line corresponds to the SDRG prediction according to the second (third) approach, 
while symbols correspond to QMC results for different chain sizes $L$. QMC error bars are smaller than symbol size.}
\end{figure}
\begin{figure}
\includegraphics[width=0.99\columnwidth]{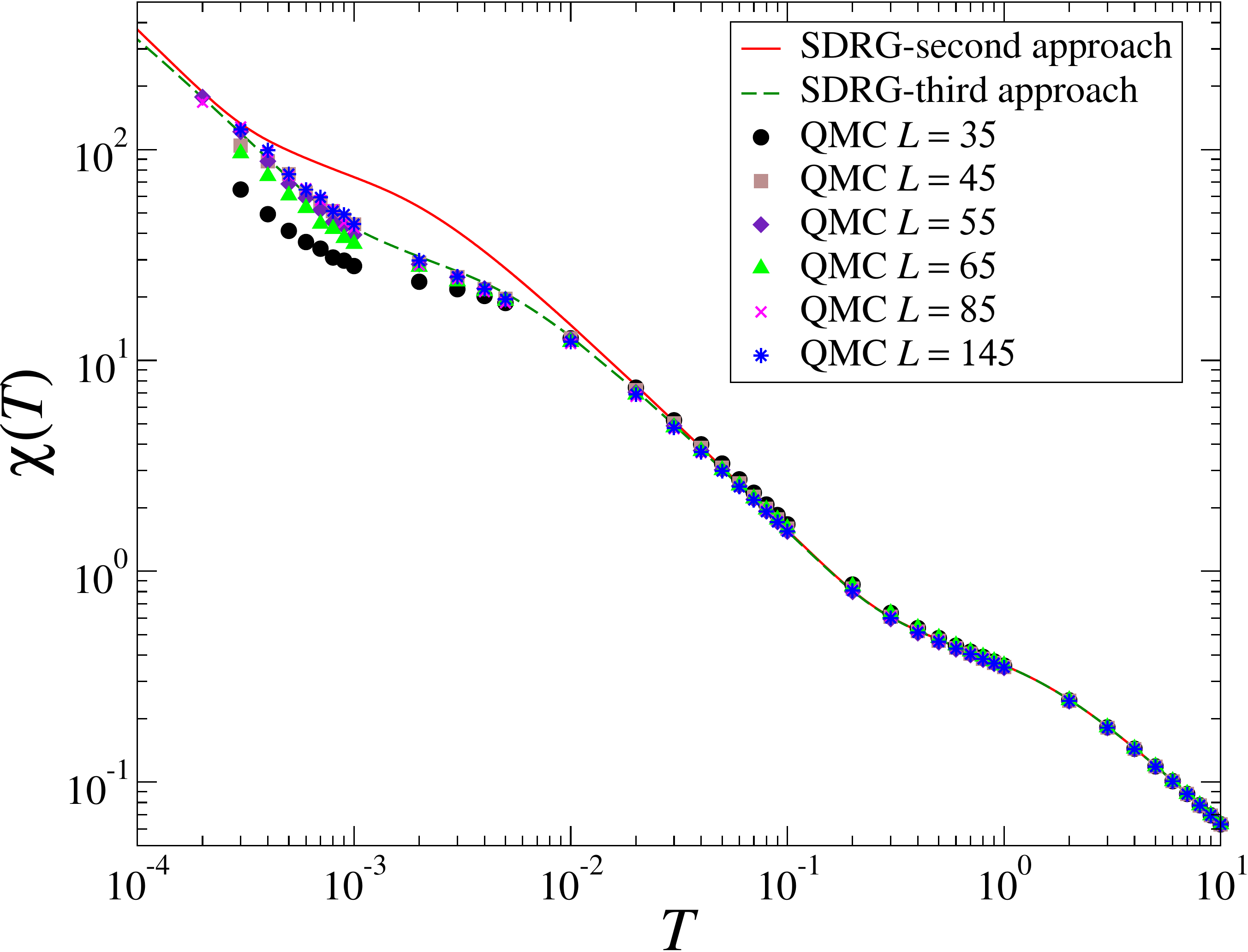} 
\caption{\label{chi-fibonacci-r20} Temperature dependence of the magnetic susceptibility
for the spin-$1$ Fibonacci-Heisenberg chain, similar as Fig.~\ref{chi-fibonacci-r10}, but with a coupling ratio $r=20$.}
\end{figure}

For the corresponding spin-$1$ chain, Figs. \ref{chi-fibonacci-r10}
and \ref{chi-fibonacci-r20} show the temperature dependence of the 
susceptibility according to the second and third SDRG approaches,
along with QMC data, for coupling ratios $r=10$ and $r=20$, respectively.
Clearly, the agreement with low-temperature numerical data is significantly 
better for the third SDRG approach, and improves as the number $L$ of spins
in the chain increases. Notice the shoulders in the susceptibility curves
(e.g., slightly to the left of $T\simeq 10^0$ and $10^{-2}$ in Fig. \ref{chi-fibonacci-r10})
at temperatures close to energy scales related to the effective $J_b$ bonds.

As the QMC calculations involve chains with an odd number of spins, the susceptibility
does not vanish at low temperatures even when the ground state is gapped. However,
for the coupling ratios used in Figs. \ref{chi-fibonacci-r10} and 
\ref{chi-fibonacci-r20}, the energy scale of the gaps, 
according to Eq. \eqref{eq:fibgap}, 
correspond to temperatures below $10^{-8}J_b$, much lower than the 
temperatures that could be reached in our simulations.

\subsection{Gap and string order correlations of the Fibonacci $S=1$ chain as a function of the coupling ratio}
\label{sec:DMRG-Fibo}

In order to check whether a sufficiently large coupling ratio could induce a gapless aperiodic 
singlet phase, we now present a numerical determination of the spin gap for different values 
of the coupling ratio and different chains lengths, using the DMRG method. 

{\it DMRG simulation details.} We simulate aperiodic $S=1$ chains with a number $L$ of spins, 
with open boundary conditions, using DMRG~\cite{white92} formulated in the matrix-product state 
formalism~\cite{McCulloch07}. We use an SU$(2)$-symmetric formulation~\cite{McCulloch02}, 
taking advantage of the symmetry of the Hamiltonian \eqref{eq:H}, which reduces considerably 
the number of states $m$ to be kept in the DMRG calculation. We nevertheless find that the 
convergence to the ground states in different total spin sector $S_T=0,1$ or $2$ is particularly 
difficult to achieve for large $L$ and large coupling ratio $r$, which we attribute to the 
aperiodicity in the system. To ensure convergence, we use a specific warming procedure where we 
increase sequentially the number $m$ of SU$(2)$ states kept, typically by values of $20$ or $50$, 
up to values of $m$ where the ground-state energy no longer varies. For the largest Fibonacci 
chains (here $L=378$), the maximum number of SU$(2)$ states was $m=1000$, 
corresponding to approximatively $4800$ U$(1)$ states. For each value of $m$ in this 
warming procedure, we perform a very large (sometimes more than $200$) number of sweeps, 
again checking that the energy does not vary.

{\it Numerical determination of gaps.} Depending on the parity of the chain size $L$, the ground-state is found to be in the $S_T=0$ 
sector (for even $L$) or the $S_T=1$ sector (odd $L$), as expected. In the Haldane phase, 
the energy difference between these two sectors is expected to decrease exponentially with 
increasing $L$ for open chains, due to the presence of spin-$\frac{1}{2}$ degrees of 
freedom near the boundaries.
\footnote{We also performed calculations for periodic boundary conditions, but convergence
was far more problematic than for open chains.}
Similar to what was done in the original DMRG study of uniform $S=1$ 
chains~\cite{white92}, we compute the gap $\Delta$ as the energy difference between this 
ground-state and the energy of the ground-state in the $S_T=2$ sector: 
$\Delta= E_0(S_T=2)-E_0(S_T=0/1)$. We simulate chains with sizes $L=14,22,35,56,90,145,234,378$ 
corresponding to the ``natural" numbers  (in the Fibonacci sequence) of bonds 
$L-1=13,21,34,55,89,144,233,377$. In the following, 
results are presented only for the specific Fibonacci bond sequence corresponding to the 
size $L$ and starting with a single letter $a$, but we checked for small $L<70$ that the same qualitative behavior is obtained when 
averaging results over the $L+1$ different possible subsequences of the Fibonacci sequence that can 
be accommodated in a chain with $L$ spins.

\begin{figure}
\includegraphics[width=0.99\columnwidth]{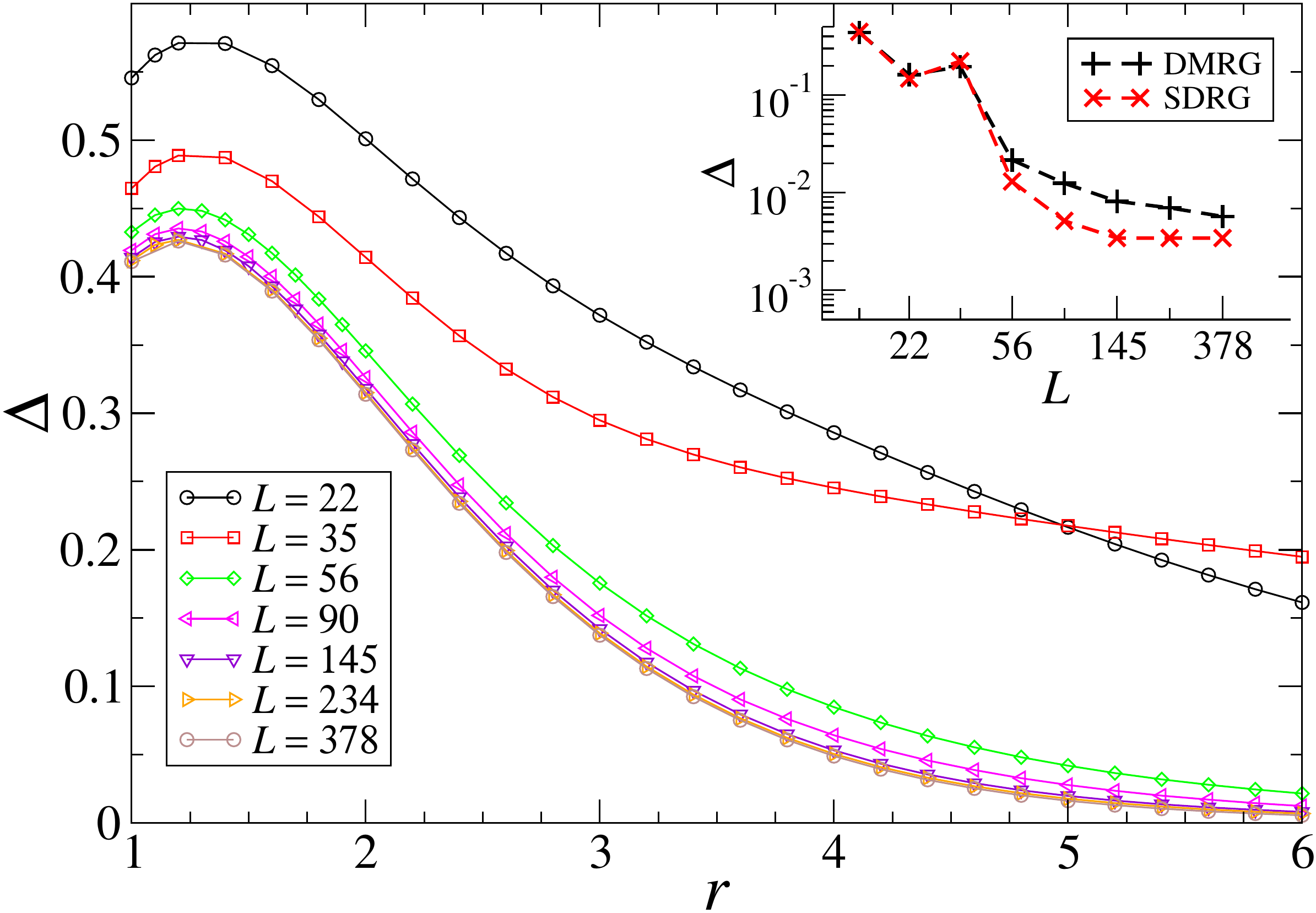}
\caption{\label{fig:gap-fibo} Spin $1$ chain modulated by the Fibonacci sequence: gap $\Delta=E_0(S_T=2)-E_0(S_T=0 / 1) $ between the lowest-lying quintuplet $S_T=2$ state energy $E_0(S_T=2)$ and the ground-state energy $E_0$ (which is either in the $S_T=0$ singlet or $S_T=1$ triplet sector, depending on the chain parity), as a function of coupling ratio $r=J_b/J_a$, for different system sizes $L$.}
\end{figure}

We present in Fig.~\ref{fig:gap-fibo} the results for the gap $\Delta$ (in units of $J_a$), 
as a function of coupling ratio $r$, for different system sizes $L$. 
It is clear from this figure that the gap does {\it not} vanish in the entire range 
$r \in [ 1, 6]$ that we simulated, even though as expected, it decreases quite considerably 
with increasing $r$. 

Notice that we should not expect the DMRG gaps 
to be direcly comparable to those predicted by Eq. \eqref{eq:fibgap}, which is valid in the
infinite-chain, large-modulation limit, and disregards boundary effects. These turn out to be
quite important, especially for the small chain lengths accessible via DMRG. Instead, we present
in the inset of Fig.~\ref{fig:gap-fibo} a comparison between the DMRG gaps for the strongest
modulation for which reliable data are available, $r=6$, and the corresponding open-chain SDRG 
predictions (see App. \ref{finite-chain-gap}). The agreement is quite good for small chains, 
but discrepancies arise
for $L\geq 56$, due to the fact that, as the effective coupling ratio decreases for increasing
system size [see Eq. \eqref{eq:fibap0rec}], the perturbative
calculations underlying the SDRG approach become less precise, leading to errors in the
gap estimate. Nevertheless, for still larger
chains ($L=234$ and $L=378$), the curves clearly approach each other. 

\begin{figure}
\includegraphics[width=0.99\columnwidth]{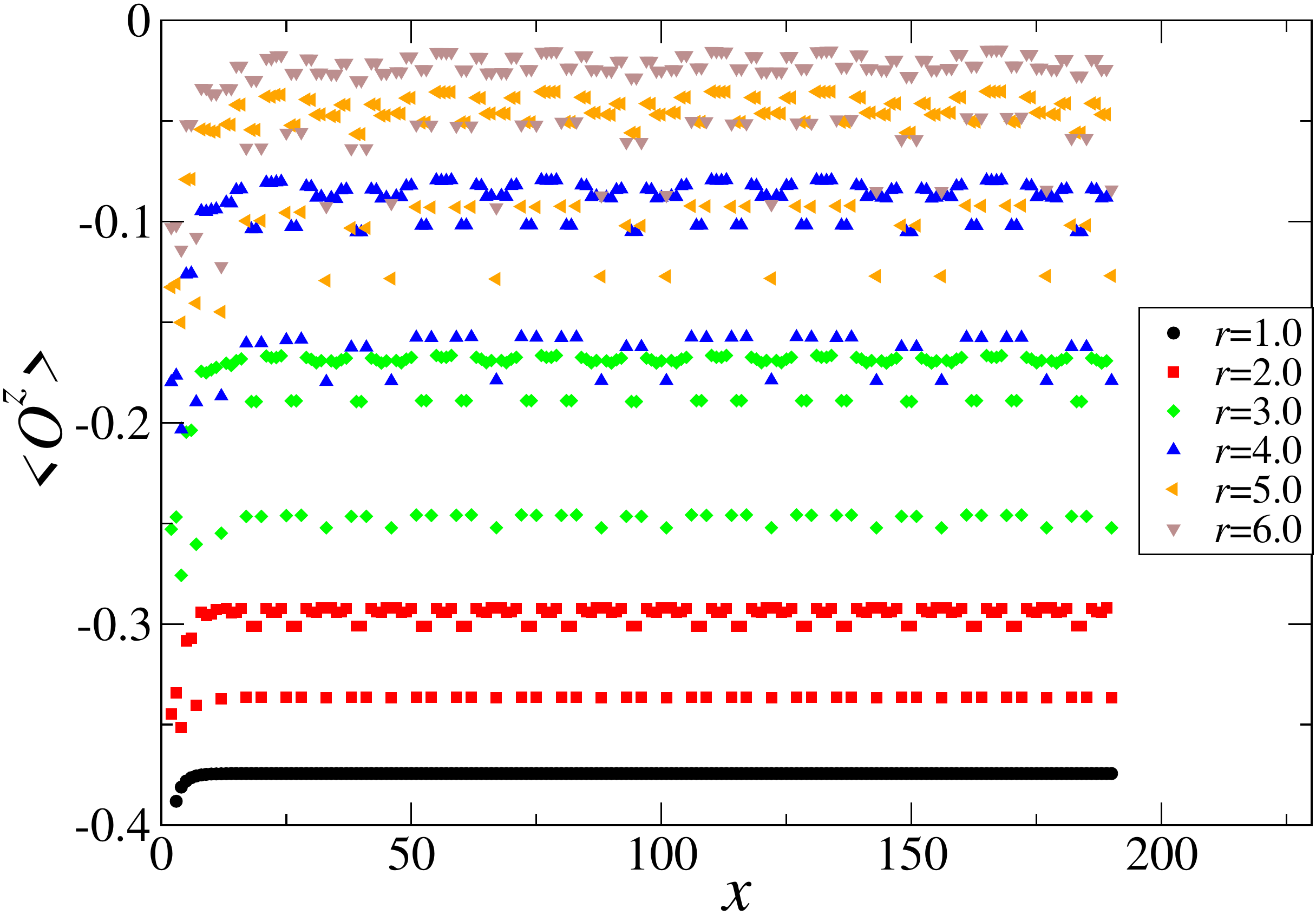}
\caption{\label{fig:string-fibo} Spin $1$ chain modulated by the Fibonacci sequence: string order correlation function $\langle O^z(x=|i-j|) \rangle$ as a function of distance $x$ taken starting from the quarter-chain point $i=L/4$ up to the maximum value $x=L/2$, for a $L=376$ chain and different aperiodicity strengths $r=J_b/J_a$.}
\end{figure}

{\it String order.} The previous gap results indicate that the Haldane phase is not destroyed by imposing a Fibonnaci aperiodic sequence for the couplings. This is furthermore confirmed by the numerical DMRG computation of the string order correlation function,~\cite{dennijs}
$$\langle O^z(i,j) \rangle = \langle S^z_i \exp( \i \pi \prod_{k=i+1}^{j-1} S^z_k ) S^z_j \rangle,$$
as a function of the distance $x=|j-i|$. The string order correlation function takes non-vanishing values in the large distance limit in the Haldane phase~\cite{dennijs} and is thus a good indicator of the continuity of the Haldane phase as the strength of the aperiodicity is increased. We represent in Fig.~\ref{fig:string-fibo} $\langle O^z(x=|j-i|) \rangle$ with $i=L/4$ and $x$ running from $0$ to $L/2$ (we consider the initial and final points to minimize effects due to the open boundary conditions) for selected values of the coupling ratio $r$, for the largest $L=378$ system simulated. A real-space correlation function such as $\langle O^z(x) \rangle$ is inevitably non-monotonous for such aperiodic systems, but the results of Fig.~\ref{fig:string-fibo} indicate that the string order does not vanish up to $r=6$, albeit it reaches smaller thermodynamic values (when $x\rightarrow \infty$) as $r$ is increased, as expected from the gap behavior.

Overall, the DMRG results on the gap and string order support the conclusion of SDRG (approaches $1$ and $3$) that the gapped Haldane phase remains robust against Fibonacci aperiodicity.

\section{\label{63-sequence}The spin-$1$ chain with couplings following the 
6-3 sequence}

We now study the effects of geometric fluctuations induced by couplings
following the 6-3 sequence on the spin-$1$ Heisenberg chain. The 6-3
sequence is defined by the substitution rule
\begin{equation}
\sigma_{63}:\left\{ \begin{array}{l}
a\to babaaa\\
b\to baa\end{array}\right.,
\label{eq:63subrule}
\end{equation}
starting from a single letter (either $a$ or $b$). The wandering exponent
characterizing pair fluctuations in the 6-3 sequence\cite{vieira05b} is 
$\omega=\ln 2/\ln 5\simeq0.43$, and thus we expect for the spin-$1\over2$
chain and for the strong-modulation spin-$1$ chain a dynamical scaling
characterized by the stretched exponential form
\begin{equation}
\Delta\left(\ell\right)\sim \exp\left(\ell/\ell_0\right)^\omega,
\end{equation}
with $r_0$ a nonuniversal constant. As described below, this is exactly
what we obtain from the SDRG scheme.

\subsection{The first approach}

\begin{figure}
\includegraphics[width=0.99\columnwidth]{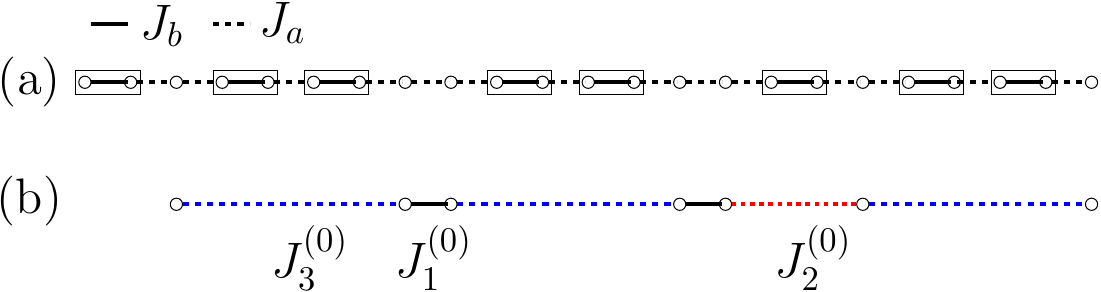} 
\caption{\label{seq63-initial-ap0}(a) Coupling distribution of the
spin-$1$ Heisenberg chain according to the 6-3 sequence.
(b) Effective chain obtained from the first SDRG approach
after a single lattice sweep.}
\end{figure}

Figure \ref{seq63-initial-ap0}(a) shows the bond distribution prescribed
by the 6-3 sequence. Assuming again $J_{b}>J_{a}$, the first SDRG lattice sweep
generates two kinds of effective bonds, exactly as in the case of the
Fibonacci-Heisenberg chain (see Fig. \ref{fibonacci-chain-ap0}). Furthermore, the 
remaining $J_a$ couplings can be reinterpreted as a third kind of effective bond,
so that we can write
\begin{equation}
J_{1}^{(0)}=J_{a},\quad
J_{2}^{(0)}=\left(\frac{4}{3}\right)\frac{J_{a}^{2}}{J_{b}},\quad
J_{3}^{(0)}=\left(\frac{4}{3}\right)^{2}\frac{J_{a}^{3}}{J_{b}^{2}},
\end{equation}
as long as $J_{b}>\sqrt{4/3}J_{a}$.

\begin{figure}
\includegraphics[width=0.99\columnwidth]{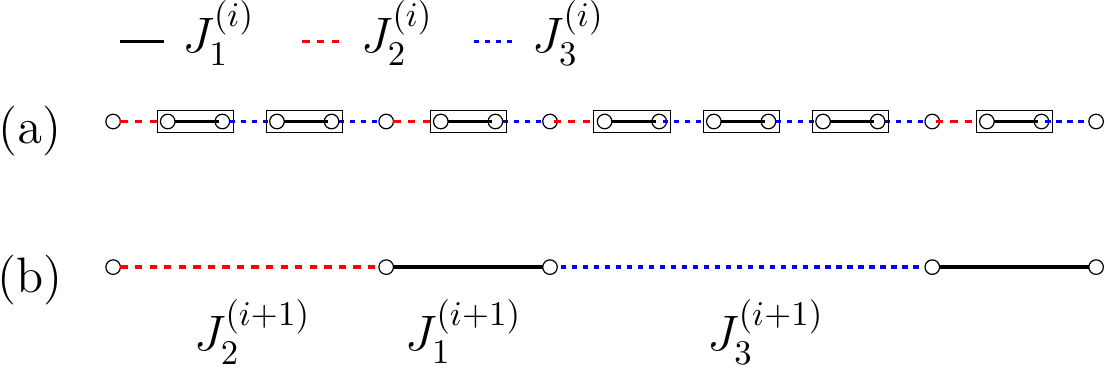} 
\caption{\label{seq63-recursive-ap0}
Self-similar coupling distribution obtained from the first SDRG approach
for subsequent lattice sweeps. Singlets are formed between spins
connected by the strong effective bonds $J_1^{(i)}$.}
\end{figure}
Subsequent SDRG lattice sweeps yield a scale-invariant coupling distribution,
as shown in Fig. \ref{seq63-recursive-ap0}, leading to
a set of recurrence equations given by
\begin{eqnarray}
J_{1}^{(j+1)} & = & \left(\frac{4}{3}\right)\frac{J_{2}^{(j)}J_{3}^{(j)}}{J_{1}^{(j)}},
\quad J_{2}^{(j+1)}=\left(\frac{4}{3}\right)^{2}\frac{J_{2}^{(j)}\left[J_{3}^{(j)}
\right]^{2}}{\left[J_{1}^{(j)}\right]^{2}},\nonumber \\
J_{3}^{(j+1)} & = & \left(\frac{4}{3}\right)^{3}\frac{J_{2}^{(j)}\left[J_{3}^{(j)}
\right]^{3}}{\left[J_{1}^{(j)}\right]^{3}},\label{recursive-seq63-ap0}
\end{eqnarray}
which are valid as long as $J_{1}^{(j)}>J_{2}^{(j)}$. This last condition is
true only for $J_b > (4/3)J_a$. 

Defining coupling ratios between the parameters $J_{1}$, $J_{2}$
and $J_{3}$, we can write the recurrence equations 
\begin{eqnarray}
\rho^{(j+1)} & \equiv & \frac{J_{1}^{(j+1)}}{J_{3}^{(j+1)}}=
\left[\frac{3}{4}\rho^{(j)}\right]^{2},\nonumber \\
\sigma^{(j+1)} & \equiv & \frac{J_{1}^{(j+1)}}{J_{2}^{(j+1)}}=
\frac{3}{4}\rho^{(j)},\label{recursive-ratio-seq63-ap0}
\end{eqnarray}
making it clear that, under the condition $J_b > (4/3)J_a$,
there is a single, infinite-modulation fixed point, 
$\rho^\infty=\sigma^\infty=\infty$. Thus, the
first SDRG approach predicts that, in the strong-modulation limit, 
couplings following the 6-3 sequence induce a gapless aperiodic singlet phase,
whose dynamical scaling form is calculated in Sec. \ref{subsec:dyn63}.
A rough estimate of the critical point separating the Haldane phase from the
gapless phase is provided by the condition $J_b > (4/3)J_a$.%, corresponding
% to a critical coupling ratio
% \begin{equation}
% r_{c}^{\mathrm{1st}}=4/3.\label{critical-point-ap0}
% \end{equation}

\subsection{The second approach}

\begin{figure}
\includegraphics[width=0.99\columnwidth]{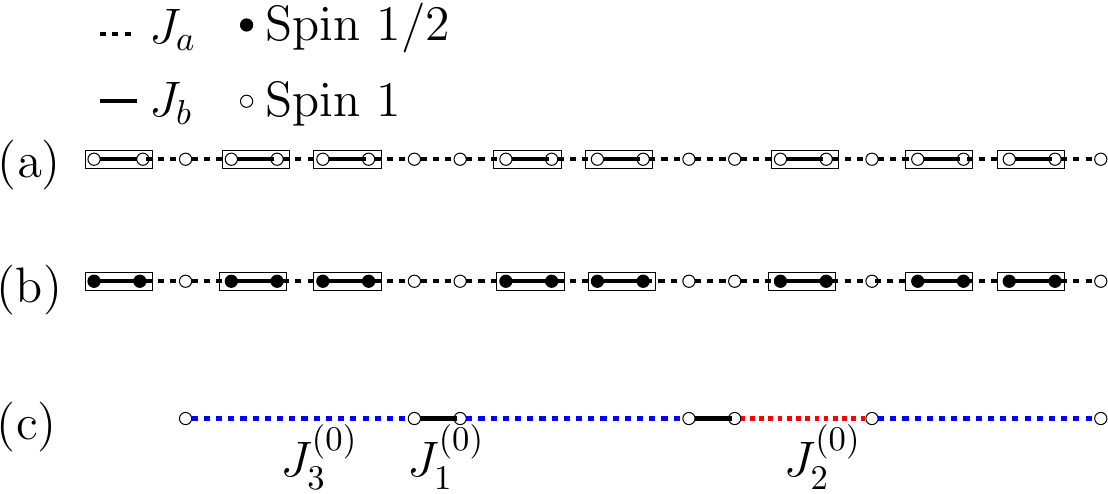} 
\caption{\label{seq63-initial-ap1}First step of renormalization of 
the spin-$1$
Heisenberg chain with couplings following the 6-3 sequence
according to the second SDRG approach. (a) The original chain.
(b) Spin-$1$ pairs connected by (strong) $J_b$ bonds are
replaced by spin-$1\over 2$ pairs. 
(c) Spin-$1\over2$ pairs are decimated, yielding effective couplings between
remaining $S=1$ spins.}
\end{figure}
Figure \ref{seq63-initial-ap1} shows the results of applying the
second SDRG approach to the spin-$1$ Heisenberg chain with couplings
modulated by the 6-3 sequence. In Fig. \ref{seq63-initial-ap1}(b),
all spin pairs connected by $J_{b}$ bonds are replaced by spin-$1\over2$
pairs after the first lattice sweep, 
and for $J_{b}\gtrsim 1.91\,J_{a}$ all 
spin-$1\over2$ pairs are then decimated, 
leading to the configuration in 
Fig. \ref{seq63-initial-ap1}(c), with three effective bonds
given by
\begin{equation}
J_{1}^{(0)}=J_{a},\quad J_{2}^{(0)}=\left(\frac{1}{2}\right)\frac{J_{a}^{2}}{J_{b}},
\quad J_{3}^{(0)}=\left(\frac{1}{2}\right)^{2}\frac{J_{a}^{3}}{J_{b}^{2}}.
\end{equation}

% 
% 
% The condition $J_{b}>2J_{a}$ sets a critical point that can be written
% in term of the ration 
% \begin{equation}
% r_{crit}^{(0)}=2.
% \end{equation}

\begin{figure}
\includegraphics[width=0.99\columnwidth]{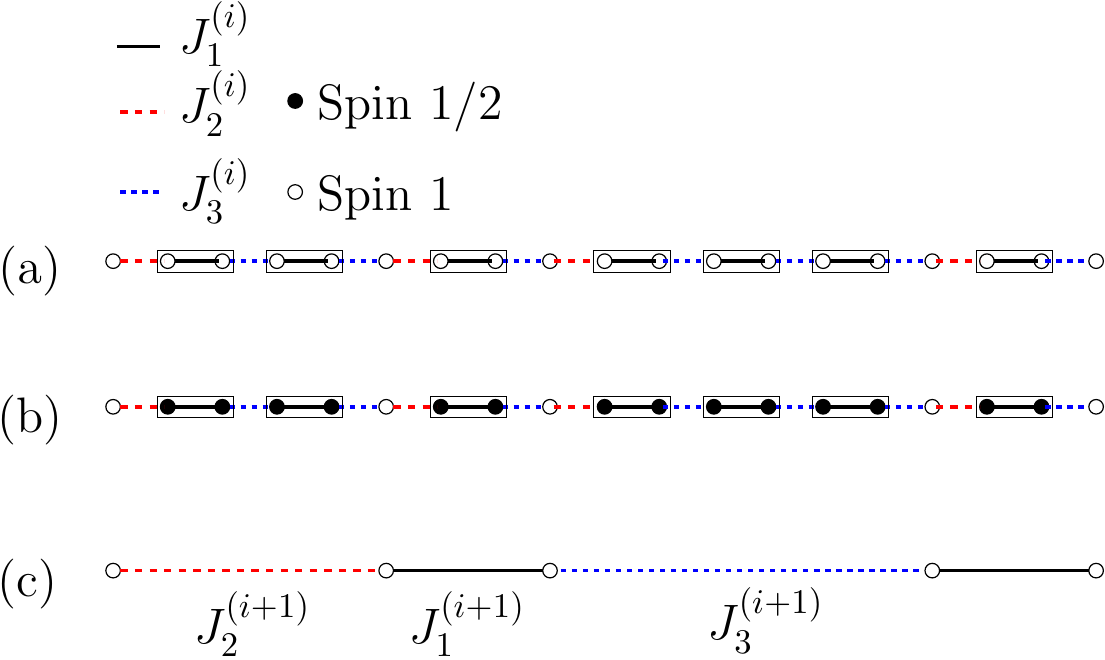} 
\caption{\label{seq63-recursive-ap1}Self-similar coupling distribution 
obtained from the second SDRG approach
for subsequent RG steps. (a) Effective chain consisting only of $S=1$ spins.
(b) Strongly connected spin-$1$ pairs form spin-$1\over2$ pairs.
(c) Spin-$1\over2$ pairs are decimated, giving rise to a new effective chain, 
again consisting only of spin-$1$ pairs, with an invariant coupling distribution.}
\end{figure}
Starting from the configuration in Fig. \ref{seq63-initial-ap1}(c),
each subsequent SDRG steps involve two consecutive sweeps through the
lattice, the first one replacing all spin-$1$ pairs connected by
$J_1$ bonds by spin-$1\over2$ pairs, which are then decimated to yield
new effective couplings. This is illustrated in Fig. 
\ref{seq63-recursive-ap1}, and leads to the recurrence relations
\begin{eqnarray}
J_{1}^{(j+1)} & = & \left(\frac{1}{2}\right)\frac{J_{2}^{(j)}J_{3}^{(j)}}{J_{1}^{(j)}},
\quad J_{2}^{(j+1)}=\left(\frac{1}{2}\right)^{2}\frac{J_{2}^{(j)}
\left[J_{3}^{(j)}\right]^{2}}{\left[J_{1}^{(j)}\right]^{2}},\nonumber \\
J_{3}^{(j+1)} & = & \left(\frac{1}{2}\right)^{3}\frac{J_{2}^{(j)}\left[J_{3}^{(j)}
\right]^{3}}{\left[J_{1}^{(j)}\right]^{3}},
\end{eqnarray}
in which $j$ labels the SDRG step. These equations are valid as long as
$J_{1}^{(j)}>\frac{3}{2}J_{2}^{(j)}$, a condition that is always verified
for $J_{b}\gtrsim 1.91\,J_{a}$.

As in the case of the first approach, we can define the coupling ratios
$\rho\equiv J_1/J_3$ and $\sigma\equiv J_1/J_2$, whose recurrence
relations read
\begin{eqnarray}
\rho^{(j+1)} & = & \left[2\rho^{(j)}\right]^{2},\nonumber\\
\sigma^{(j+1)} & = & 2\rho^{(j)}.\label{recursive-seq63-ap1}
\end{eqnarray}
These also point to an infinite-modulation fixed point,
$\rho^\infty=\sigma^\infty=\infty$, so that predictions
from the first and the second approach are now in qualitative
agreement, although, as for the Fibonacci-Heisenberg chain,
the energy levels predicted by the two approaches (estimated from
the effective $J_1$ bonds) are distinct.

If $J_{b}\lesssim 1.91\,J_{a}$, numerical implementations of the second SDRG approach
(not detailed here) predict the renormalization of a different set
of bonds than in the first SDRG step, according to the recipe associating
the energy scale with the bond clusters yielding the largest local gap. 
However, for $1.69\,J_a\lesssim J_b\lesssim 1.91\,J_a$, the distribution of effective
couplings in Fig. \ref{seq63-recursive-ap1}(a) is eventually reached, so
that the scheme still predicts a gapless, aperiodic singlet phase as
the ground state. For $J_b\lesssim 1.69\,J_a$, however, the distribution
of effective couplings arrives at a dimerized spin-$1\over2$ chain,
a state equivalent to the Haldane phase. Thus, within the approximations
leading to the second SDRG approach, $J_b\simeq 1.69\,J_a$ corresponds to the
critical point separating a gapped from an aperiodicity-dominated
gapless phase.

\subsection{The third approach}

\begin{figure}
\includegraphics[width=0.99\columnwidth]{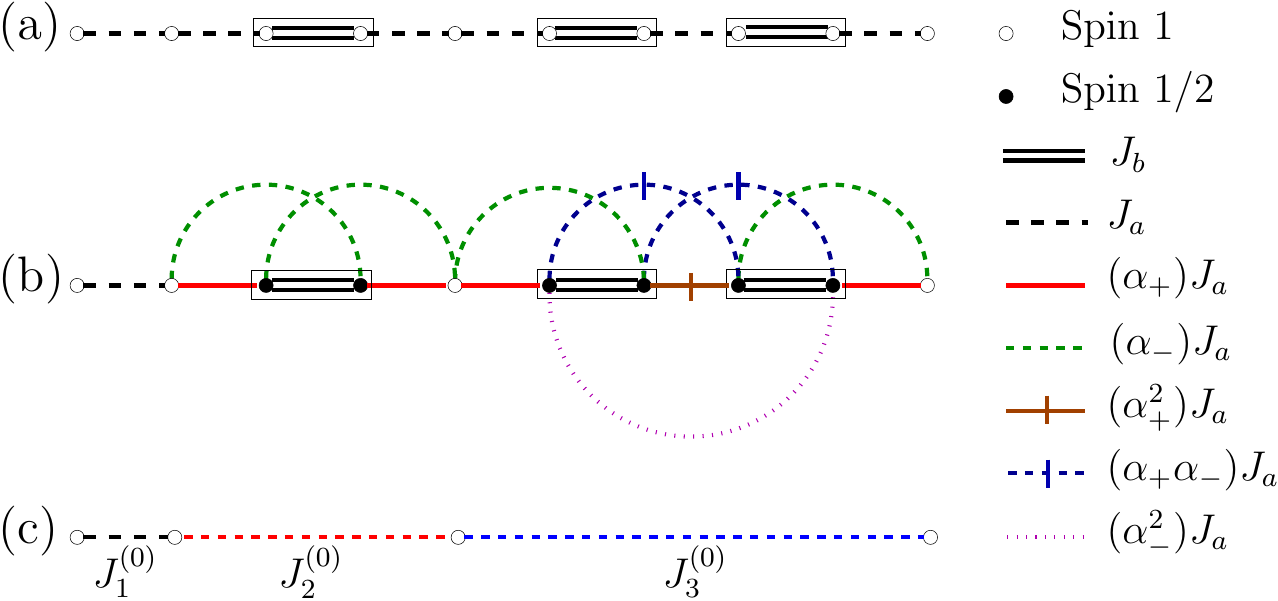}
\caption{\label{seq63-initial-ap2}
First step of the renormalization of 
the spin-$1$ Heisenberg chain with couplings following the
6-3 sequence,
according to the third SDRG approach. (a) The original chain.
(b) Spin-$1$ pairs connected by (strong) $J_b$ bonds are
replaced by spin-$1\over 2$ pairs, are further-neighbor couplings are
produced. 
(c) Spin-$1\over2$ pairs are decimated, yielding effective couplings between
remaining $S=1$ spins.}
\end{figure}
Figure \ref{seq63-initial-ap2} shows the first step of the renormalization
of the spin-$1$ Heisenberg chain with couplings following the
6-3 sequence, according to the third SDRG approach. Forming spin-$1\over2$
pairs from strongly connected spin-$1$ pairs, and assuming
$J_b>3J_a$, second- and third-order perturbation theory
leads to effective couplings which, along with the remaining $J_a$
bonds, define a set of effective bonds
\begin{eqnarray}
J_1^{(0)} & = & J_a,\nonumber \\
J_2^{(0)} & = & \frac{(\alpha_+-\alpha_-)^2}{2}
\frac{J_a^2}{J_b}=\left(\frac{4}{3}\right)\frac{J_a^2}{J_b},\nonumber \\
J_3^{(0)} & = & \frac{(\alpha_+-\alpha_-)^4}{4}
\frac{J_a^3}{J_b^2}=\left(\frac{4}{3}\right)^2
\frac{J_a^3}{J_b^2},\label{recursive-seq63-ap2}
\end{eqnarray}
exactly as in the first SDRG approach. Notice that, as in the 
Fibonacci-Heisenberg chain, further-neighbor couplings are
introduced in the middle of the RG step, but eliminated at the end
for strong enough modulation. Nevertheless, they are essential in
obtaining from the third approach the same effective couplings
predicted by the first approach.
\footnote{Of course, for weak or moderate
modulation further-neighbor couplings extend over larger and
larger distances, making numerical implementations of the SDRG
quite intricate. These cases are not considered here.}

% The critical point, which is the condition between the initial parameter
% to generate the asymptotic configuration is given by 
% \begin{equation}
% r_{crit}^{(0)}=2\alpha_{+}.
% \end{equation}

\begin{figure*}
\includegraphics[width=0.8\textwidth]{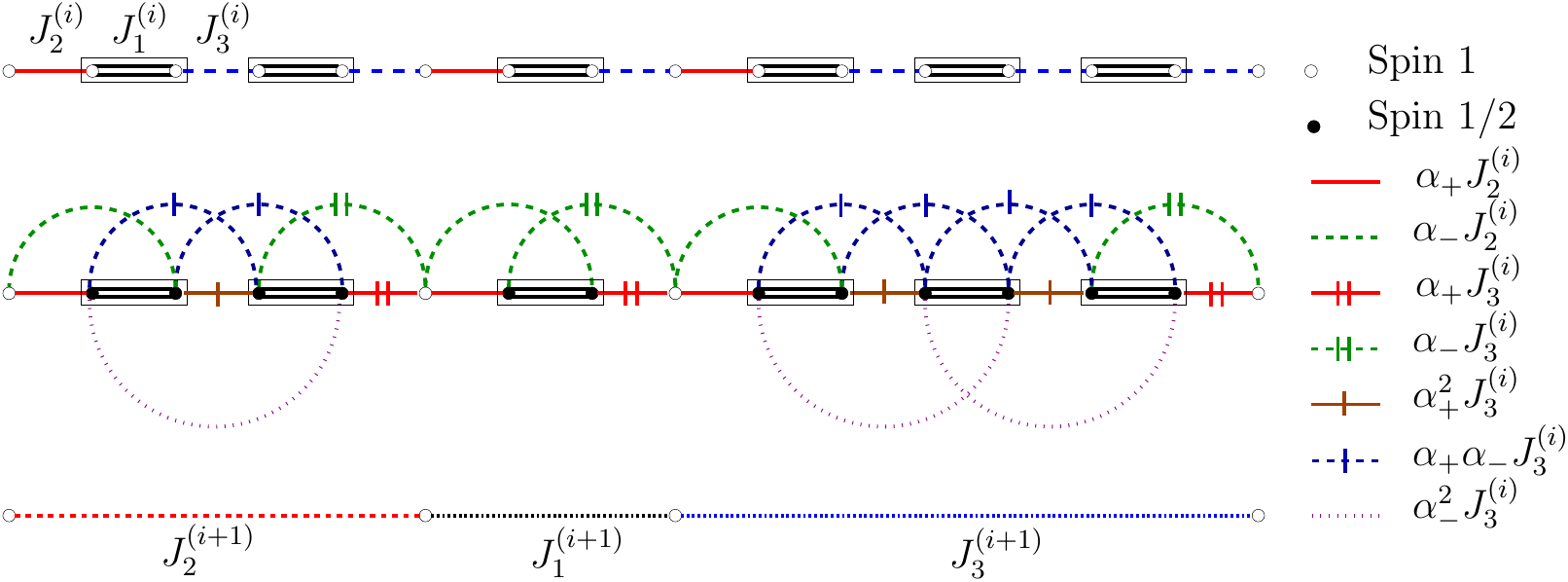}
\caption{\label{seq63-recursive-ap2}
Self-similar coupling distribution 
obtained from the third SDRG approach for subsequent RG steps. 
(a) Effective chain consisting only of $S=1$ spins.
(b) Strongly-connected spin-$1$ pairs form spin-$1\over2$ pairs, and 
further-neighbor bonds are formed.
(c) Spin-$1\over2$ pairs are decimated, giving rise to a new effective chain, 
again consisting only of spin-$1$ pairs, with an invariant coupling distribution.
}
\end{figure*}
Subsequent SDRG steps start from the coupling distribution in
Fig. \ref{seq63-initial-ap2}(c), shown in expanded form in Fig.
\ref{seq63-recursive-ap2}(a). After forming spin-$1\over2$
pairs from spin-$1$ pairs coupled by effective $J_1$ bonds, these
spin-$1\over2$ pairs are decimated, taking into account the
presence of further neighbor bonds, to yield, from second-,
third- and fourth-order perturbation theory, new effective
couplings obeying the recurrence relations
\begin{eqnarray}
J_{1}^{(j+1)} & = & \frac{(\alpha_{+}-\alpha_{-})^{2}}{2}
\frac{J_{2}^{(j)}J_{3}^{(j)}}{J_{1}^{(j)}}
=\left(\frac{4}{3}\right)\frac{J_{2}^{(j)}J_{3}^{(j)}}{J_{1}^{(j)}},\nonumber \\
J_{2}^{(j+1)} & = & \frac{(\alpha_{+}-\alpha_{-})^{4}}{4}
\frac{J_{2}^{(j)}\left[J_{3}^{(j)}\right]^{2}}{\left[J_{1}^{(j)}\right]^{2}}=
\left(\frac{4}{3}\right)^2
\frac{J_{2}^{(j)}\left[J_{3}^{(j)}\right]^{2}}{\left[J_{1}^{(j)}\right]^{2}},
\nonumber \\
J_{3}^{(j+1)} & = & \frac{(\alpha_{+}-\alpha_{-})^{6}}{8}
\frac{J_{2}^{(j)}\left[J_{3}^{(j)}\right]^{3}}{\left[J_{1}^{(j)}\right]^{3}}=
\left(\frac{4}{3}\right)^3
\frac{J_{2}^{(j)}\left[J_{3}^{(j)}\right]^{3}}{\left[J_{1}^{(j)}\right]^{3}},
\nonumber \\
\label{recursive-seq63}
\end{eqnarray}
valid as long as $J_b>\alpha_+^2Ja$.

Thus, for sufficiently strong modulation, the first and the third SDRG approaches
yield the same quantitative predictions for the ground-state properties and
the energy levels, while the second approach qualitatively agrees with
the other two.

In the presence of moderate or weak modulation, in which the perturbative 
calculations underlying the SDRG scheme become increasingly inadequate,
quantitative predictions are expected to depend on finer details of
the first and third approaches. Indeed, for $\alpha_+^2 J_a<J_b<3J_a$,
numerical implementations of the third approach still predict a 
gapless ground state, although with a slightly different set of renormalized
bonds in the first RG step. 
However, for $J_b\lesssim \alpha_+^2 J_a$, the third approach eventually
leads to an effective chain composed of spin-$1\over2$ objects with a dimerized
distribution of effective couplings, thus predicting a gapped phase.
Of course, for such a range of coupling ratios, we do not expect any 
of the predictions for the critical coupling ratio to be precise.

% \subsection{General results.}
% 
% Now we can write the general form results for the Heisenberg chain
% modulated by the 6-3 sequence in the same way we have done for the
% Fibonacci sequence. The recursive equations are
% \begin{eqnarray}
% J_{1}^{(j+1)} & = & \gamma\frac{J_{2}^{(j)}J_{3}^{(j)}}{J_{1}^{(j)}},\hspace{0.3cm}J_{2}^{(j+1)}=\gamma^{2}\frac{J_{2}^{(j)}\left[J_{3}^{(j)}\right]^{2}}{\left[J_{1}^{(j)}\right]^{2}},\nonumber \\
% J_{3}^{(j+1)} & = & \gamma^{3}\frac{J_{2}^{(j)}\left[J_{3}^{(j)}\right]^{3}}{\left[J_{1}^{(j)}\right]^{3}},\label{recursive-seq63}
% \end{eqnarray}
% and the recursive equations for the ratio are
% \begin{eqnarray}
% \rho^{(j+1)}=\left(\frac{\rho^{(j)}}{\gamma}\right)^{2},\hspace{0.5cm}\sigma^{(j+1)}=\frac{\rho^{(j)}}{\gamma}.\label{recursive-ratio-seq63}
% \end{eqnarray}
% 
% 
% One again the prefactor has the value $\gamma=1/2$ for the approach
% 1 and $\gamma=4/3$ for the approaches 0 and 2. But now due the quadratic
% dependence in the $\rho^{(j)}$ recursive equation both ratios always
% increase above the critical point. We remark that each approach predicts
% a different critical point.
% 
% These equations also hold for the spin 1/2 chain \cite{vieira05a,vieira05b}
% with $\gamma=1/2$, which is also equal to the approach 1, where the
% typical configuration for the spin 1/2 chain is equal to the configurations
% shown in the figures \ref{seq63-initial-ap0} and \ref{seq63-recursive-ap0}.
\subsection{Dynamic scaling relation}\label{subsec:dyn63}

In the strong-modulation gapless phase we can derive the dynamic
scaling relation between energy and length scales. It is natural
to assume that, as the various energy levels are estimated from
the values of the strongest effective couplings at each step
of the SDRG scheme, the relevant length
scales are the corresponding effective lengths. From the recurrence
relations in Eqs. \eqref{recursive-seq63}, and by looking at Fig.
\ref{seq63-recursive-ap2}, it can be seen that the lengths of the
effective couplings satisfy recurrence relations that can be written
in matrix form as
\begin{eqnarray}
\left[\begin{array}{cc}
\ell_{1}^{(j+1)}\\
\ell_{2}^{(j+1)}\\
\ell_{3}^{(j+1)}
\end{array}\right]=\left[\begin{array}{ccc}
1 & 1 & 1\\
2 & 1 & 2\\
3 & 1 & 3
\end{array}\right]\cdot\left[\begin{array}{ccc}
\ell_{1}^{(j)}\\
\ell_{2}^{(j)}\\
\ell_{3}^{(j)}
\end{array}\right],
\end{eqnarray}
in which again $j$ labels the SDRG steps. The matrix appearing in
the above equation has eigenvalues $\lambda_1=\lambda_2=0$ and $\lambda_3=5$,
so that, in the asymptotic limit, all effective lengths scale as
\begin{equation}
\ell^{(j)}\sim\lambda_{3}^{j}.\label{characteristic-length}
\end{equation}

The energy levels, being proportional to the value of the largest bond in each
iteration,  scale as $\Delta_{j}\sim J_{1}^{(j)}$. Thus, 
by solving the recurrence relations in Eqs. \eqref{recursive-ratio-seq63-ap0} 
and \eqref{recursive-seq63}, we can write 
% \begin{equation}
% J_{1}^{(j+1)}=\frac{4}{3}\left[\rho^{(j)}\right]^{-\frac{3}{2}}J_{1}^{(j)}.
% \end{equation}
% 
% 
% For the ratio $\rho^{(j)}$ we can iterate the equation \eqref{recursive-ratio-seq63}
% to obtain
% \begin{equation}
% \rho^{(j)}=\exp\left[2^{j}\ln\rho^{(0)}-2\ln\gamma\left(2^{j}-1\right)\right],
% \end{equation}
% and using this result in the recursive equation for the singlet bond
% we find
% \begin{equation}
% J_{1}^{(j)}=\gamma^{j}\exp\left[-\frac{3}{2}\ln\left(\rho^{(0)}\gamma^{-2}\right)\left(2^{j}-1\right)-3j\ln\gamma\right].
% \end{equation}
% 
% 
% Finally using the last result with the equation \eqref{characteristic-length}
% to eliminate the $j$ dependence we obtain the dynamical scaling given
% by
\begin{equation}
\Delta_{j}\sim\ell_{j}^{\frac{-2\ln(4/3)}{\ln 5}}\exp\left[-\frac{3}{2}
\ln\left(\frac{9}{16}\rho^{(0)}\right)\ell_{j}^{\omega}\right],
\end{equation}
with $\omega=\frac{\ln2}{\ln5}$. 
As expected, apart from unimportant constants, this is the same 
stretched-exponential form 
obeyed by the spin-$1\over2$ Heisenberg chain with couplings following the 6-3
sequence.

\subsection{Comparison with QMC simulations}

Using the same independent-singlet approximation described for the
Fibonacci-Heisenberg chain in Sec. \ref{sec:FibQMC}, we can obtain
SDRG predictions for the susceptibility at zero field when couplings
follow the 6-3 sequence. Only a small adaptation is necessary,
as the self-similar coupling distribution is distinct from the 
6-3 sequence itself. Thus, we must take into account that the fraction 
of $J_b$ bonds in the original chain is $f_{b}=\frac{1}{3}$, while the
fraction of $J_{1}$ bonds in the self-similar distribution is 
$f_{J_{1}}=\frac{2}{5}$. Below, the results of the independent-singlet
approximation are compared
with quantum Monte Carlo simulations.

\begin{figure}[htm]
\includegraphics[width=0.99\columnwidth]{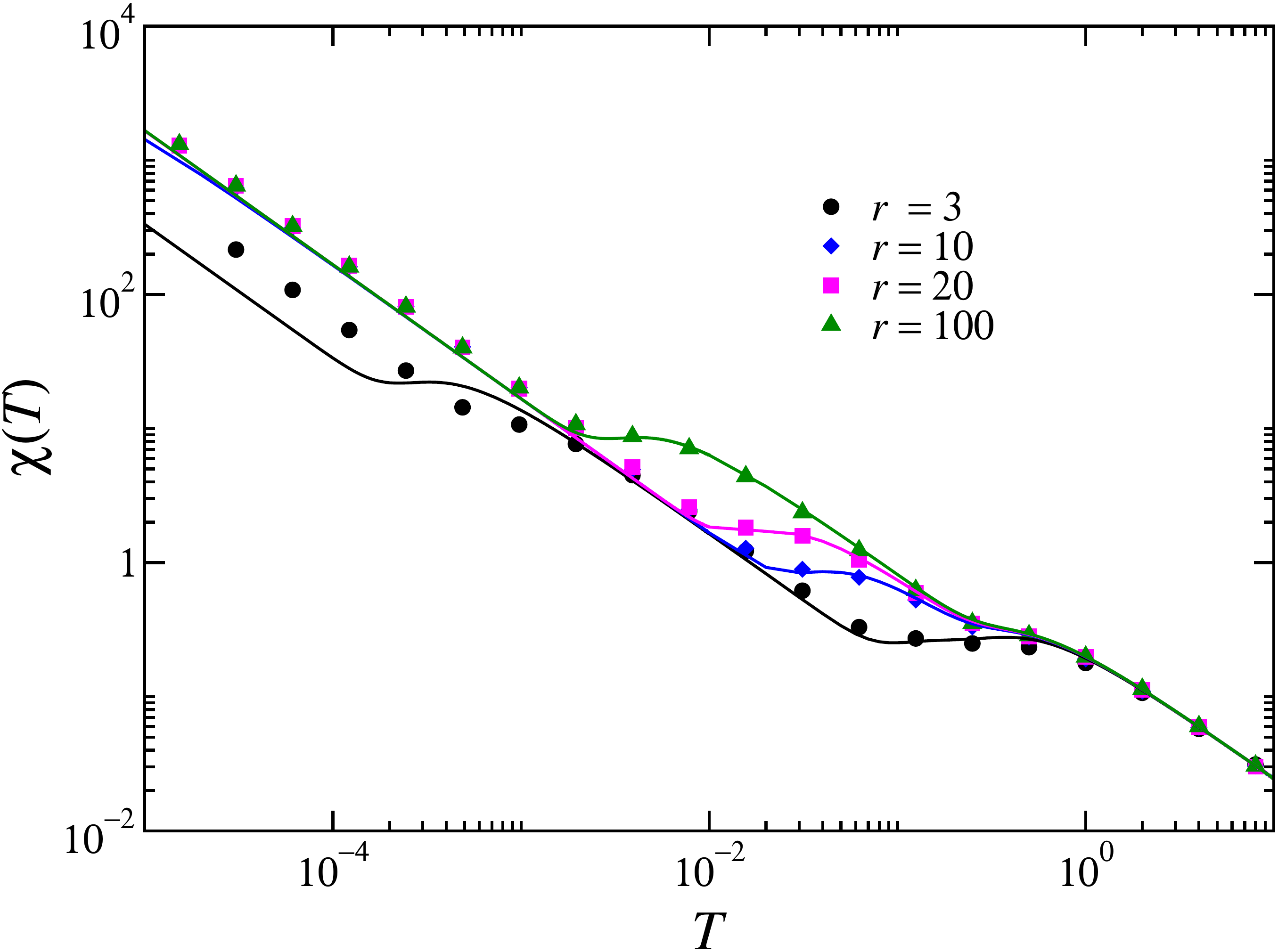}
\caption{\label{chi_of_T_seq63_L75_OBC_S1.2}
Temperature dependence of the magnetic susceptibility for the spin-$1\over2$ Heisenberg chain with aperiodic couplings following the 
6-3 sequence. Solid lines are the SDRG predictions for various coupling ratios
$r=J_b/J_a$, while symbols indicate the corresponding QMC results obtained for $L=75$ sites.}
\end{figure}
For the spin-$1\over2$ chain with $L=75$ sites the results are shown in Fig.
\ref{chi_of_T_seq63_L75_OBC_S1.2}. As expected, the agreement between
the SDRG prediction and QMC simulations is better for larger 
coupling ratios $r= J_b/J_a$.

\begin{figure}
\includegraphics[width=0.99\columnwidth]{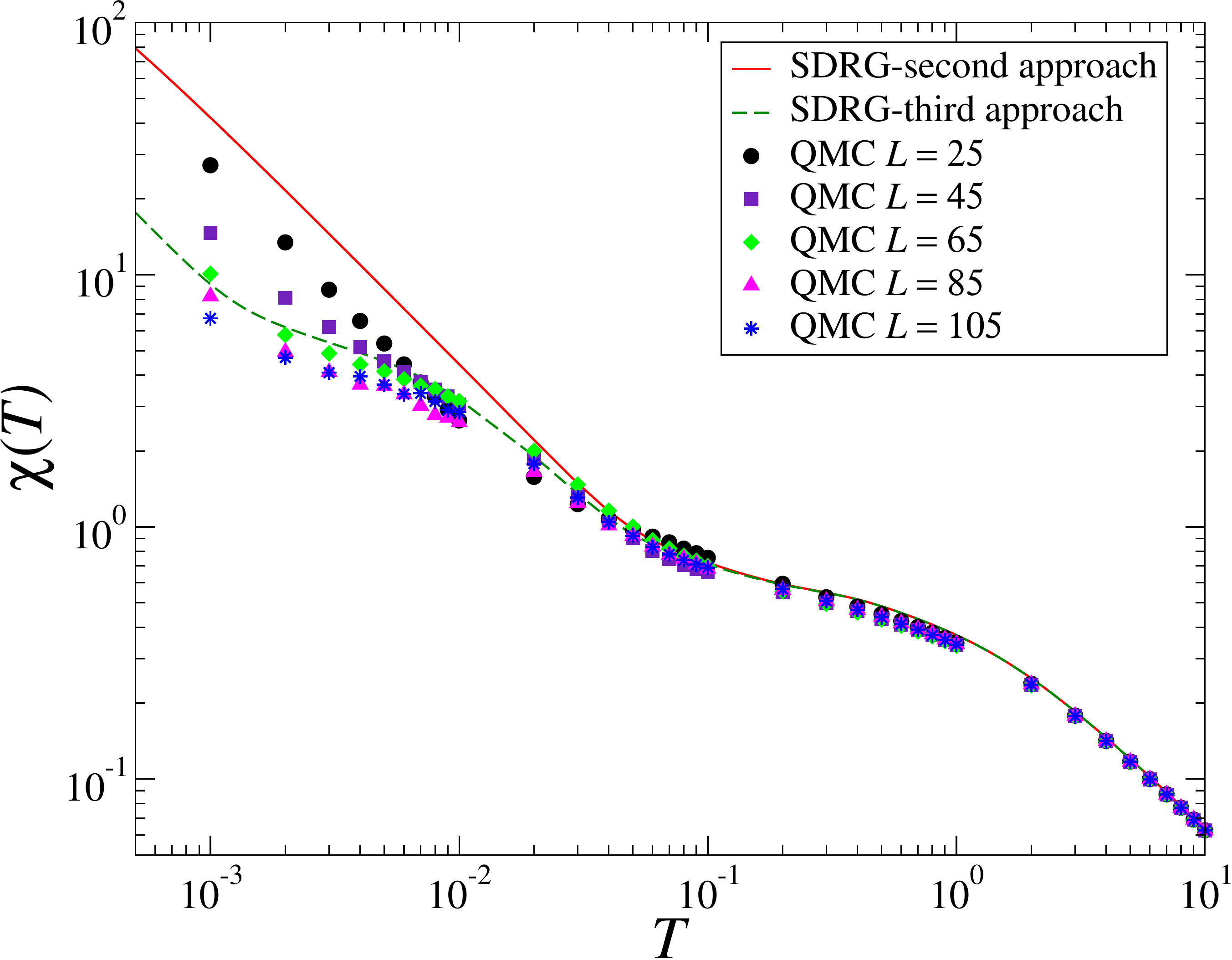} 
\caption{\label{chi-seq63-r5}
Temperature dependence of the magnetic susceptibility for the spin-$1$ Heisenberg chain with aperiodic couplings following the 
6-3 sequence with coupling ratio $r=5$. Solid (dashed) 
line corresponds to the SDRG prediction according to the second (third) approach, 
while symbols correspond to QMC results for different chain sizes $L$. QMC error bars are smaller than symbol size.}
\end{figure}
\begin{figure}
\includegraphics[width=0.99\columnwidth]{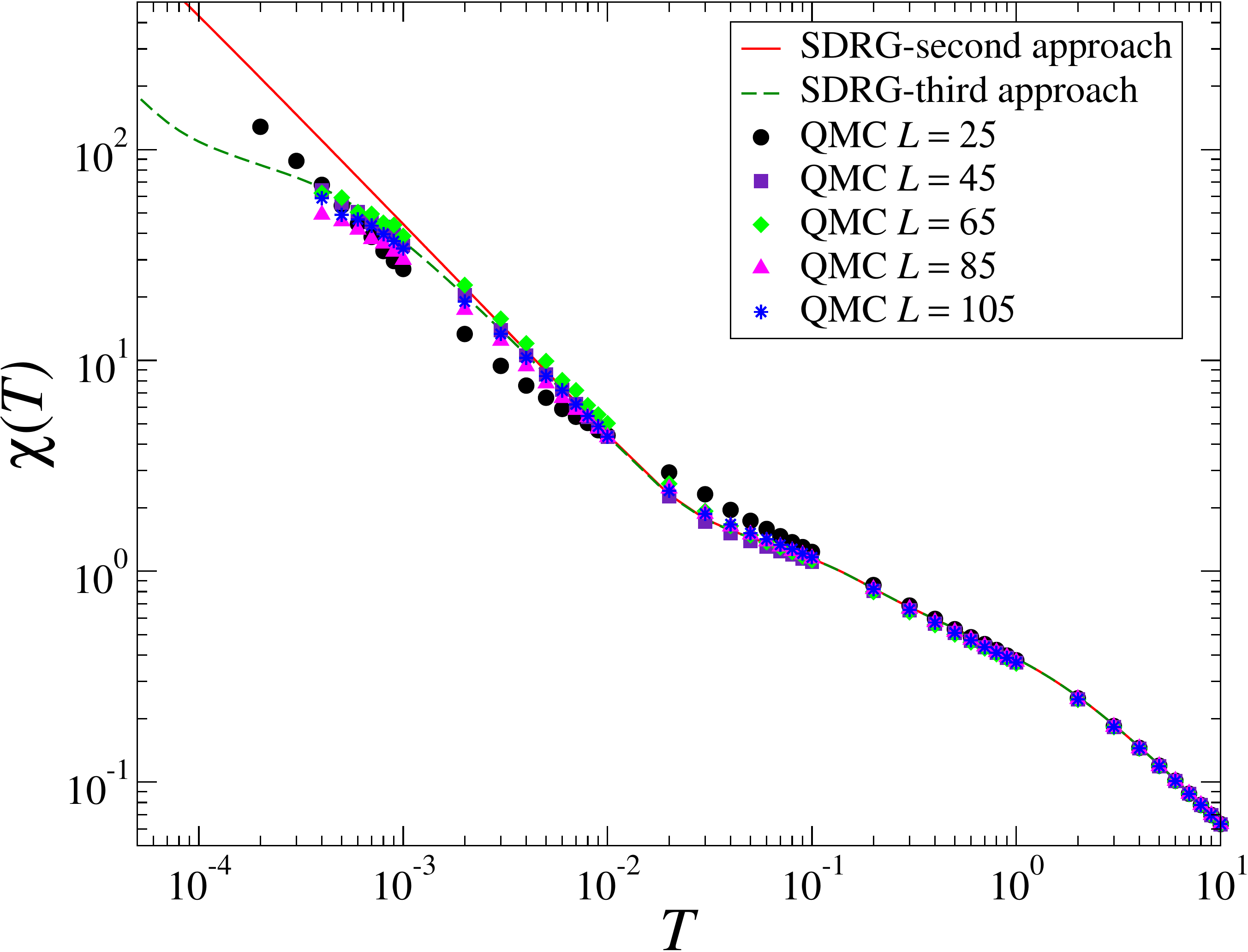} 
\caption{\label{chi-seq63-r10} Temperature dependence of the magnetic susceptibility
for the spin-$1$ Heisenberg chain with aperiodic couplings following the 
6-3 sequence, similar as Fig.~\ref{chi-seq63-r5}, but with a coupling ratio $r=10$.}
\end{figure}
In Figs. \ref{chi-seq63-r5} and \ref{chi-seq63-r10} we show
the results for the spin-$1$ chain with coupling ratios $r=5$ and $r=10$ respectively.
As in the case of the Fibonacci-Heisenberg chain, it is clear that the QMC results are in
better agreement with the predictions of the third SDRG approach.
Again notice the shoulders in the susceptibility curves
close to temperatures corresponding to energy scales related to the effective $J_b$ bonds.

\subsection{Gap and string order of the 6-3 $S=1$ chain as a function of the coupling ratio}

\begin{figure}
\includegraphics[width=0.99\columnwidth]{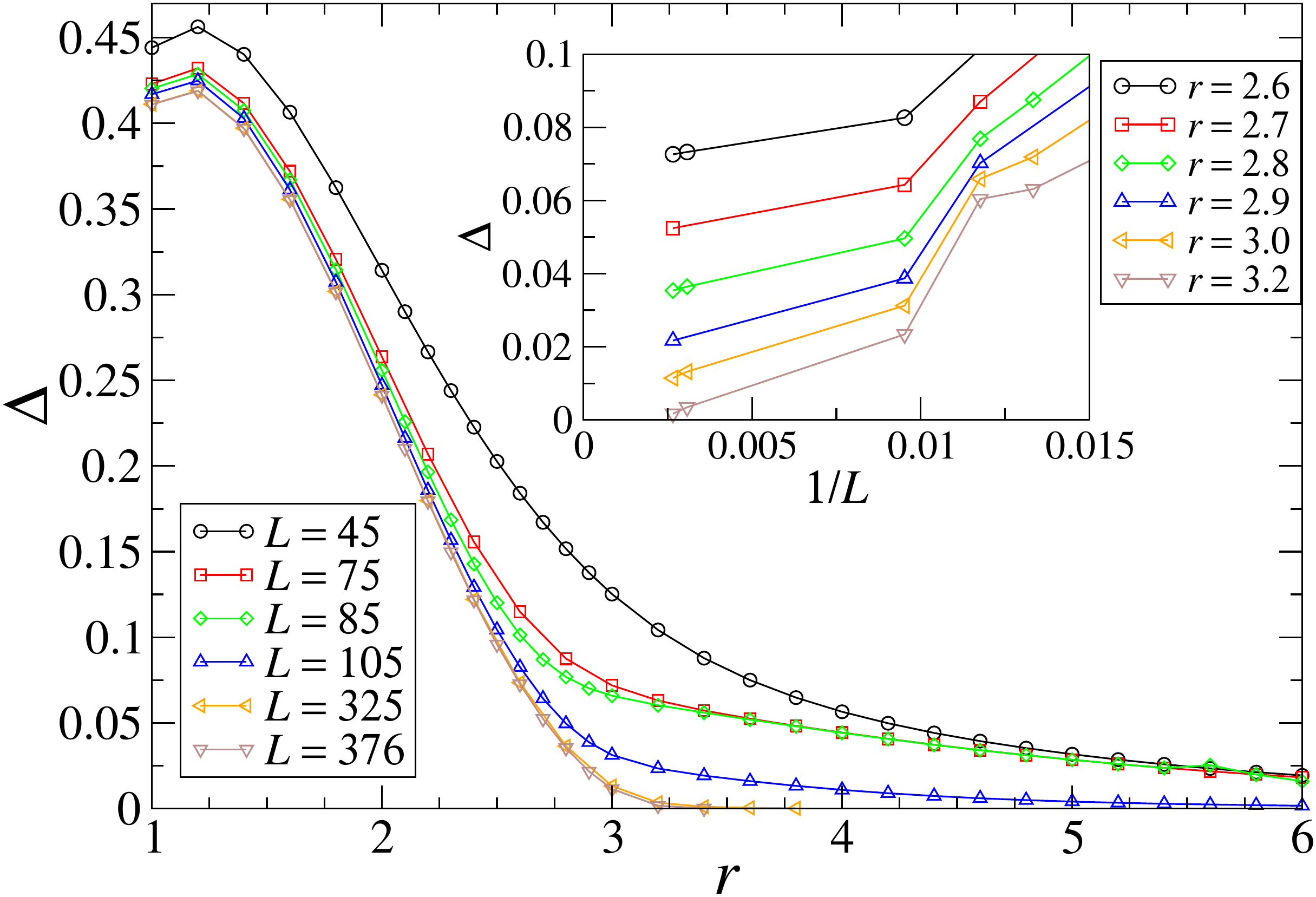}
\caption{\label{fig:gap-63} Spin-$1$ chain modulated by the 6-3 sequence: 
gap $\Delta=E_0(S_T=2)-E_0(S_T=0\/1) $  between the lowest-lying quintuplet $S_T=2$ 
state energy $E_0(S_T=2)$ and the ground-state energy $E_0$ (which is either in the 
$S_T=0$ singlet or $S_T=1$ triplet sector, depending on the chain parity), as a function 
of coupling ratio $r=J_b/J_a$, for different system sizes $L$.}
\end{figure}

We use the same DMRG procedure described in Sec.~\ref{sec:DMRG-Fibo} to compute the spin 
gap $\Delta$ (in units of $J_a$) for open $S=1$ spin chains modulated by the 6-3 sequence. 
We use systems of sizes $L=45,75,85,105,325,376$ and display the results in 
Fig.~\ref{fig:gap-63}. Our calculations reveal that the spin gap $\Delta$ clearly vanishes 
for sufficiently large systems, when the coupling ratio $r$ is large enough. 
For the largest systems considered ($L=325$ and $L=376$), our simulations did not converge 
for too large values of $r \geq 4$, but the behavior at smaller $r$ and for smaller 
$L$ clearly indicates that the gap must vanish for these cases.

These results are again in agreement with the SDRG calculations, which indicate 
that the Haldane gap must vanish above a critical modulation $r_c$, to give rise to the 
gapless aperiodic singlet phase. The critical value $r_c$ at which this quantum phase transition 
takes place is difficult to estimate precisely due to strong finite-size effects arising from a 
small gap. Considering the largest system available, we can nevertheless ascertain that 
the system is gapless at  $r=3.4$. The inset of Fig.~\ref{fig:gap-63} displays the spin 
gap $\Delta$ as a function of inverse system size $1/L$, for values of the coupling ratio 
close to the transition. We can tentatively deduce a value $r_c \approx 2.9(2)$, even 
though this phenomenological determination has to be taken with care. 
Even though the SDRG prediction $r_c \simeq 1.73$ (from approach $3$) is different,
it is subject to a large uncertainty, since for such small values of the bond ratio
the perturbative calculations become much imprecise, and 
we can nevertheless conclude that the numerical calculations of the spin gap support
the SDRG prediction of a gapless phase at large enough (but finite) value of the coupling 
ratio $r$. 

\begin{figure}
\includegraphics[width=0.99\columnwidth]{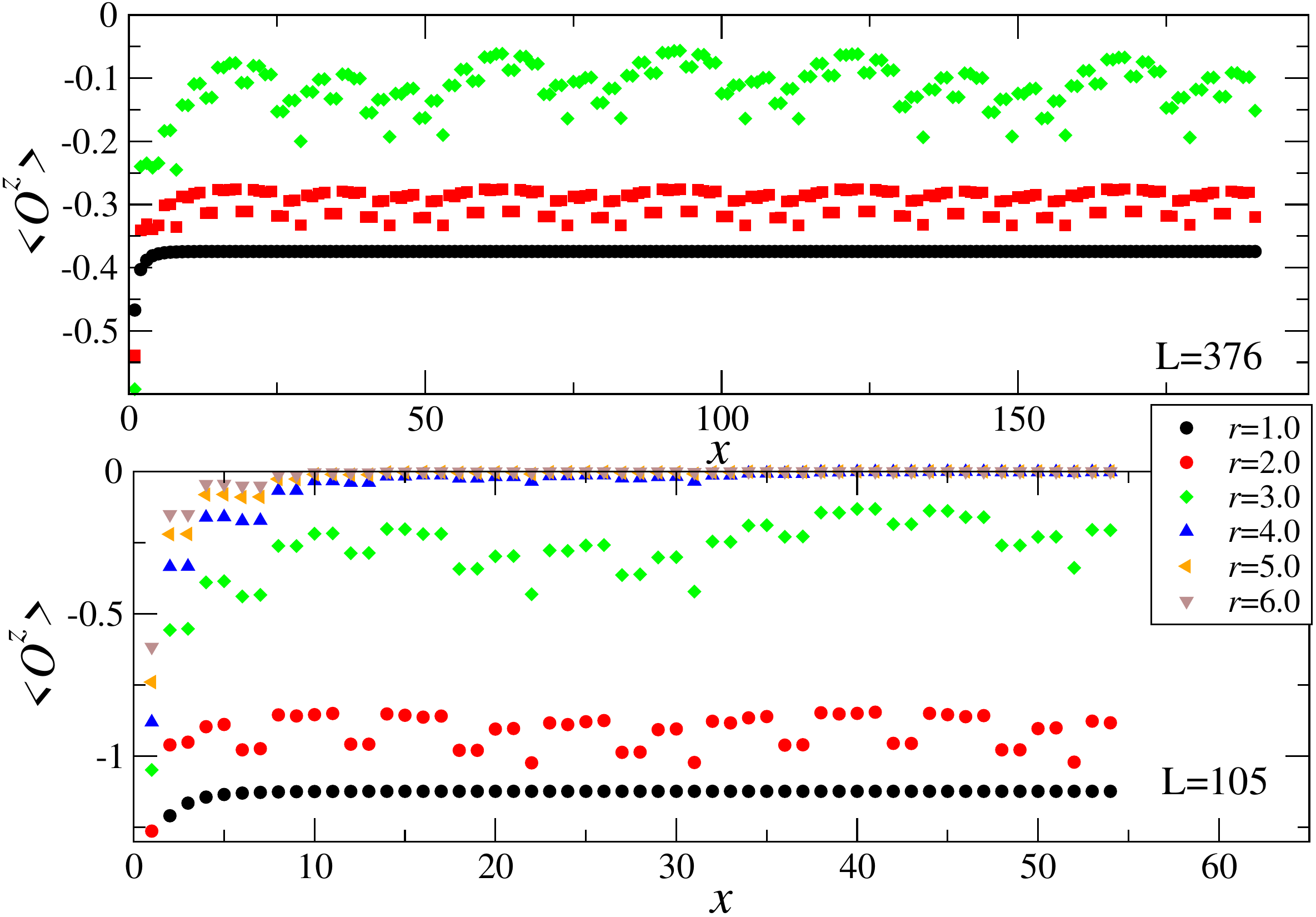}
\caption{\label{fig:string-63} Spin-$1$ chain modulated by the 6-3 sequence:  string order correlation function $\langle O^z(x=|i-j|) \rangle$ as a function of distance $x$ taken starting from the quarter-chain point $i=L/4$ up to the maximum value $x=L/2$, for a $L=376$ chain (top) and a $L=105$ chain (bottom) and different aperiodicity strengths $r=J_b/J_a$.}
\end{figure}

We finally confirm this finding by computing the string order correlation function $\langle O^z(x=|j-i|) \rangle$ using the same setup as presented in Sec.~\ref{sec:DMRG-Fibo}. We present in Fig.~\ref{fig:string-63} the results of our simulations for the maximum chain size $L=376$ where we could reach convergence for $r=1,2,3$ (top panel) and for a smaller chain size $L=105$ where convergence was ensured up to $r=6$ (bottom panel), for integer values of $r$. For $L=376$, the long-distance behavior of  $\langle O^z(x) \rangle$ indicates that the Haldane phase is still present in this finite-size sample up to $r=3$, albeit with a  small string order parameter for this latter value of $r$,  in agreement with the small gap value found for this system. On the other hand, for $r=4,5,6$, results for the smaller sample $L=105$ already clearly indicate that the string order vanishes in the long-distance limit, nicely confirming that the Haldane phase has disappeared. Due to the irregular behavior in $x$, we did not 
attempt to perform finite-size scaling on the string-order correlation function for different system sizes to estimate the critical coupling value $r_c$, but our results for the largest sample $L=376$ are consistent with the estimate $r_c=2.9(2)$ obtained from the gap estimate.

\section{\label{Conc}Discussion and Conclusions}

In this paper, we investigated the effects of aperiodic but deterministic bond modulations 
on the zero and low-temperature properties of the spin-$1$ Heisenberg chain. We presented explicit
results for aperiodic bonds generated by two different binary substitution rules, associated with
the Fibonacci and the 6-3 sequences. 

For the Fibonacci-Heisenberg chain, whose geometric fluctuations are gauged by a wandering 
exponent $\omega=0$, calculations based on different adaptations of the SDRG scheme yielded
conflicting results. While the SDRG approach of Monthus \emph{et al.}~\cite{monthus97,monthus98},
which allows for the appearance of effective $S={1\over2}$ spins as the transformation proceeds,
predicts that for strong bond modulation the ground state corresponds to a gapless,
aperiodicity-dominated phase, the inclusion of nonfrustrating next-nearest-neighbor effective bonds
in the SDRG scheme points to a gapped ground state, and to the stability of the Haldane phase
towards any finite Fibonacci modulation. This is the same prediction as obtained from the simplest SDRG
scheme which only involves $S=1$ spins, and is supported by both quantum Monte Carlo and
DMRG calculations.

On the other hand, for the Heisenberg chains with bonds following the 6-3 sequence, characterized
by stronger geometric fluctuations ($\omega\simeq0.43$), all SDRG approaches give the same
qualitative prediction, according to which the Haldane phase should be stable in the presence of
weak bond modulation (as measured by the ratio $r$ between the strong and weak bonds 
$J_b$ and $J_a$), while strong bond modulation ($r\gg 1$) drives the ground state towards a gapless
aperiodicity-dominated phase, similar to the one obtained for the analogous $S={1\over2}$ Heisenberg chain.
Again, this prediction is nicely supported by quantum Monte Carlo and DMRG calculations. 

Although we only presented explicit calculations for two aperiodic sequences, we can draw more 
general conclusions for the strong-modulation regime based on known results for
the $S={1\over2}$ Heisenberg chain.~\cite{vieira05b} For virtually all binary substitution sequences
characterized by a wandering exponent $\omega\geq 0$,
it is possible to write recursion relations for a main bond ratio in the form~\cite{vieira05b}
\begin{equation}
 r^{(j+1)}=\gamma\left[r^{(j)}\right]^k,
\end{equation}
where $\gamma$ is a constant, $r^{(j)}$ is the bond ratio calculated at the $j$th iteration of 
the SDRG transformation, 
and $k$ is an integer related to the wandering exponent $\omega$ and to the rescaling factor
$\tau$ of the transformation by
\begin{equation}
 \omega = \frac{\ln k}{\ln\tau}.
\end{equation}

While for the SDRG approach of Monthus {\it et al}. the constant $\gamma$ is greater than $1$, the
other approaches predict $0<\gamma<1$. For $k\geq 2$ ($\omega>0$) the effective bond
ratio diverges along the iterations, as long as the bare bond ratio $r^{(0)}$ is large enough,
irrespective of the value of $\gamma$, thus always driving the system towards a gapless
phase in the strong-modulation limit. On the other hand, for $k=1$ ($\omega=0$), the constant
$\gamma$ defines whether the flow of the effective bond ratio is directed towards the Haldane
phase of unit effective bond ratio or to the opposing aperiodicity-dominated phase. Therefore,
we predict that only for sequences for which the wandering exponent is zero the different
SDRG approaches will offer conflicting qualitative results.

In general, the presence of aperiodic bonds characterized by a positive wandering exponent
will lead to a phase transition between the Haldane phase and a gapless phase as the bond
modulation increases. In contrast to the random-bond spin-$1$ chain, however,
we do not expect an intermediate phase associated with Griffiths singularities. This is due
to the fact that the inflation symmetry of substitution sequences precludes the appearance
of arbitrarily large regions in which the system is locally in the opposite phase as
compared to the infinite chain. This is in agreement with the behavior of the aperiodic
quantum Ising chain\cite{oliveira2012} and also, in the context of nonequilibrium transitions
to an absorbing state, of the aperiodic contact process.~\cite{barghathi2013}
Furthermore, due to the fact that the critical point corresponds to a bare bond ratio of
order unity, estimates of the critical exponents of the transition from the (perturbative) 
SDRG scheme are both technically quite difficult and unreliable. Any calculations of such
quantities by numerical methods are left for future work.

\section*{Acknowledgments}
We thank J. A. Hoyos for insightful discussions. We warmly thank I. McCulloch for providing 
access to his code\footnote{See http://physics.uq.edu.au/people/ianmcc/mptoolkit/} used to 
perform the DMRG calculations. QMC calculations were partly performed using the SSE code~\cite{Alet05,ALPS} from the ALPS project~\footnote{See http://alps.comp-phys.org}. This work was performed using HPC resources from GENCI (grants x2013050225 and x2014050225) and CALMIP (grants 2013--P0677 and 2014--P0677)
and is supported by the French ANR program ANR-11-IS04-005-01, by the Brazilian agencies
FAPESP(2009/08171-3 and 2012/02287-2), CNPq (530093/2011-8 and 304736/2012-0), and FAPESB/PRONEX,
and by Universidade de S\~ao Paulo (NAP/FCx).

\appendix

\section{\label{pairomega}The wandering exponent for letter pairs}

For the Fibonacci sequence, the substitution rule for letter pairs is built by
applying three times the substitution rule $\sigma_{\textrm{fb}}$ of Eq. \ref{eq:fibsubrule},
yielding
\[
\sigma_{\textrm{fb}}^{3}:\left\{ \begin{array}{l}
a\to abaab\\
b\to aba\end{array}\right..\]
Noting that the pair $bb$ does not occur in the sequence, it follows that
\[
\sigma_{\textrm{fb}}^{(2)}:\left\{ \begin{array}{l}
aa\to(ab)(aa)(ba)(ba)(ab)\\
ab\to(ab)(aa)(ba)(ba)\\
ba\to(ab)(aa)(ba)(ab)\end{array}\right..\]

For a general pair inflation rule $\sigma^{(2)}$, an
associated substitution matrix can be defined as\[
\mathbf{M}^{(2)}=\left(\begin{array}{cccc}
m_{aa}(w_{aa}) & m_{aa}(w_{ab}) & m_{aa}(w_{ba}) & m_{aa}(w_{bb})\\
m_{ab}(w_{aa}) & m_{ab}(w_{ab}) & m_{ab}(w_{ba}) & m_{ab}(w_{bb})\\
m_{ba}(w_{aa}) & m_{ba}(w_{ab}) & m_{ba}(w_{ba}) & m_{ba}(w_{bb})\\
m_{bb}(w_{aa}) & m_{bb}(w_{ab}) & m_{bb}(w_{ba}) & m_{bb}(w_{bb})\end{array}\right),\]
where $m_{\alpha\beta}(w_{\gamma\delta})$ denotes the number of pairs $\alpha\beta$ in
the word associated with the pair $\gamma\delta$. The leading eigenvalues $\lambda_{1}$
and $\lambda_{2}$ of $\mathbf{M}^{(2)}$ define a wandering
exponent\[
\omega=\frac{\ln\left|\lambda_{2}\right|}{\ln\lambda_{1}},\]
which governs the fluctuations of the letter pairs (see Ref. \onlinecite{vieira05b} and references
therein).

\section{\label{block-del}Local gaps}

At moderate modulation, it is important to identify which
spin blocks lead to the largest local energy gap, since these
are the blocks to be renormalized according to the SDRG prescription.
In the following tables, we list the local gaps corresponding to the various
blocks produced by the aperiodic sequences used in this paper.
Table \ref{table-del-ap0} lists the blocks relevant for the first SDRG approach,
while Tab. \ref{table-del-ap1} is relevant for the second and third approaches.
The last column in each table shows the renormalized blocks, a single straight
line corresponding to an effective coupling between the spins neighboring the
original block. Additional effective couplings may appear; see Figs. \ref{rule2}
to \ref{rule4prime}.

\begin{table}[htm]
\caption{\label{table-del-ap0}Local gaps $\Delta$, in units of the bond $J$ connecting
spins in each of the various blocks relevant for the first approach. The last
column shows the corresponding renormalized block.}
\begin{ruledtabular}
\begin{tabular}{cccc}
n (block size)  & configuration  & $\Delta/J$ (gap) & renorm. block \tabularnewline
\hline 
2  & $\circ$\textemdash{}$\circ$  & 1.0 & \textemdash{} \tabularnewline
3  & $\circ$\textemdash{}$\circ$\textemdash{}$\circ$  & 1.0 & $\circ$ \tabularnewline
4  & $\circ$\textemdash{}$\circ$\textemdash{}$\circ$\textemdash{}$\circ$  & 0.5092
& \textemdash{} \tabularnewline
\hline 
notation:  & $\circ=\textrm{spin 1}$  & \tabularnewline
\end{tabular}\end{ruledtabular} 
\end{table}

\begin{table}[htm]
\caption{\label{table-del-ap1}Local gaps $\Delta$, in units of the bond $J$ connecting
spins in each of the various blocks relevant for the second and third approaches. The last
column shows the corresponding renormalized block, with spins connected by a bond $J^\prime$.}
\begin{ruledtabular}
\begin{tabular}{cccc}
n (block size)  & configuration  & $\Delta/\left|J\right|$ (gap) & renorm. block \tabularnewline
\hline 
2  & $\circ$\textemdash{}$\circ$  & 3.0 &$\bullet$\textemdash{}$\bullet$ ($J^\prime>0$)\tabularnewline
2  & $\bullet$\textemdash{}$\circ$  & 1.5 & $\bullet$ \tabularnewline
2  & $\bullet$\textemdash{}$\bullet$ $(J>0)$  & 1.0 & \textemdash{} ($J^\prime>0$) \tabularnewline
2  & $\bullet$\textemdash{}$\bullet$ $(J<0)$  & 1.0 & $\circ$ \tabularnewline
3  & $\circ$\textemdash{}$\circ$\textemdash{}$\circ$  & 2.0 
   & $\bullet$\textemdash{}$\bullet$ ($J^\prime<0$)\tabularnewline
3  & $\bullet$\textemdash{}$\circ$\textemdash{}$\bullet$  & 1.0 & \textemdash{} ($J^\prime>0$) \tabularnewline
3  & $\circ$\textemdash{}$\circ$\textemdash{}$\bullet$  & 1.5 & $\bullet$ \tabularnewline
3  & $\bullet$\textemdash{}$\bullet$\textemdash{}$\bullet$  & 1.0 & $\bullet$ \tabularnewline
4  & $\circ$\textemdash{}$\circ$\textemdash{}$\circ$\textemdash{}$\circ$ & 1.8545
   & $\bullet$\textemdash{}$\bullet$ ($J^\prime>0$) \tabularnewline
4  & $\bullet$\textemdash{}$\circ$\textemdash{}$\circ$\textemdash{}$\bullet$  & 1.9142
   & $\bullet$\textemdash{}$\bullet$ ($J^\prime>0$) \tabularnewline
4  & $\circ$\textemdash{}$\circ$\textemdash{}$\circ$\textemdash{}$\bullet$  & 1.0778
   & $\bullet$ \tabularnewline
\hline 
notation:  & $\circ=\textrm{spin 1}$; $\bullet=\textrm{spin 1/2}$  & \tabularnewline
\end{tabular}\end{ruledtabular} 
\end{table}

\section{\label{third-order-pt}Third-order perturbative calculations
of effective couplings in the third SDRG approach}

We consider, as a perturbation over the local Hamiltonian $h_0$ in Eq. \eqref{h0-twopairs},
the Hamiltonian
\begin{eqnarray}
h_{1}^{\mathrm{exact}} &=& J_{a}\mathbf{S}_{4}\cdot\left(\alpha_{+}\mathbf{s}_{5}+
\alpha_{-}\mathbf{s}_{6}\right)+\nonumber \\
&+& J_{a}\mathbf{s}_{5}\cdot\left(\alpha_{+}\alpha_{-}\mathbf{s}_{7}+\alpha_{-}^{2}
\mathbf{s}_{8}\right)+\nonumber \\
&+& J_{a}\mathbf{s}_{6}\cdot\left(\alpha_{+}^{2}\mathbf{s}_{7}+\alpha_{+}\alpha_{-}
\mathbf{s}_{8}\right)+\nonumber \\
&+& J_{a}\left(\alpha_{-}\mathbf{s}_{7}+\alpha_{+}\mathbf{s}_{8}\right)
\cdot\mathbf{S}_{9},
\end{eqnarray}
which includes both nearest- and next-nearest bonds to the spins in $h_0$.

Following degenerate perturbation theory, we find that first- and second-order corrections 
to $h_0$ are identically zero, while the third-order corrections 
arise from the eigenvalues of the matrix
\begin{eqnarray}
h^{\mathrm{eff}} & = & \sum\limits _{i\neq0,j\neq0}
\frac{\left\langle\Psi_{0}\left|h_{1}^{\mathrm{exact}}\right|\Psi_{i}\right\rangle
\left\langle\Psi_{i}\left|h_{1}^{\mathrm{exact}}\right|\Psi_{j}\right\rangle
\left\langle\Psi_{j}\left|h_{1}^{\mathrm{exact}}\right|\Psi_{0}\right\rangle}
{(E_{i}-E_{0})(E_{j}-E_{0})}+\nonumber \\
 & - & \left\langle\Psi_{0}\left|h_{1}^{\mathrm{exact}}\right|\Psi_{0}\right\rangle
 \sum\limits _{i\neq0}\frac{\left|\left\langle\Psi_{0}\left|h_{1}^{\mathrm{exact}}\right|
 \Psi_{i}\right\rangle\right|^{2}}{(E_{i}-E_{0})^{2}},
\end{eqnarray}
in which the states are obtained from direct products of the eigenstates of the spin pairs
5-6 and 7-8. Those states are: the ground state
\begin{equation}
 |\Psi_0\rangle=|\Phi_{0}\rangle_{56}\otimes|\Phi_{0}\rangle_{78},
\end{equation}
formed by combining both pairs in the singlet states defined in Eq. \eqref{singleto},
and excited states $|\Psi_i\rangle$ which are formed by singlet-triplet or triplet-triplet
combinations of the states $|\Phi_{0}\rangle$, $|\Phi_{1}^+\rangle$, $|\Phi_{1}^0\rangle$,
and $|\Phi_{1}^-\rangle$; see again Eq. \eqref{tripleto}.

Expanding the summations we arrive at an effective bond between spins 4 and 9 given
by Eq. \eqref{eq:fibjaprime}.

\section{\label{fibonacci-appendix}Fractions of letters in an infinite aperiodic sequence}

Let us consider a general two-letter substitution rule
\begin{equation}
\sigma:\left\{ \begin{array}{l}
a\to w_a\\
b\to w_b\end{array}\right.,
\end{equation}
in which $w_a$ and $w_b$ are words formed by arbitrary combinations of letters $a$ and $b$. 
If the numbers of letters $a$ and $b$ are respectively $n_a$ and $n_b$, 
after applying the substitution rule these numbers change to $n_a^\prime$ and $n_b^\prime$, 
such that
\begin{equation}
\left[\begin{array}{c}n_a^\prime \\ n_b^\prime \end{array}\right]=
\left[\begin{array}{cc}m_{aa} & m_{ab} \\ m_{ba} & m_{bb} \end{array}\right]
\left[\begin{array}{c}n_a \\ n_b \end{array}\right],
\end{equation}
$m_{\alpha\beta}$ being the number of letters $\alpha$ in the word $w_\beta$.

After many iterations of the substitution rule, assuming the convergence 
of the fractions of letters $a$ and $b$, $f_a=n_a/(n_a+n_b)$ and
$f_b=n_b/(n_a+n_b)$, it follows from the above matrix equation that we can write
\begin{equation}
 f_b=\frac{m_{ba}+f_b\left(m_{bb}-m_{ba}\right)}
 {m_{aa}+m_{ba}+f_b\left(m_{ab}+m_{bb}-m_{aa}-m_{ba}\right)},
\end{equation}
with $f_a=1-f_b$.

For the Fibonacci sequence, whose substitution rule is given by Eq. 
\eqref{eq:fibsubrule}, we have $m_{aa}=m_{ab}=m_{ba}=1$ and $m_{bb}=0$, so
that we obtain $f_b=(3-\sqrt{5})/2$. For the 6-3 sequence, with the
substitution rule in Eq. \eqref{eq:63subrule}, we have $m_{aa}=4$, $m_{ab}=m_{ba}=2$
and $m_{bb}=1$, so that $f_b=\frac{1}{3}$.

\section{\label{finite-chain-gap}Finite-chain SDRG gaps for the Fibonacci-Heisenberg chain}

\begin{figure}
\includegraphics[width=0.65\columnwidth]{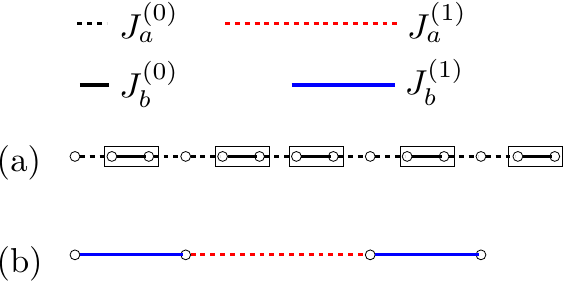} 
\caption{\label{fibonacci-chain-ap0-l14}
SDRG approach as applied to the spin-$1$ Fibonacci-Heisenberg chain
with $L=14$ sites. After sweeping over the effective chain in
(b), all spins are eliminated.}
\end{figure}
\begin{figure}
\includegraphics[width=\columnwidth]{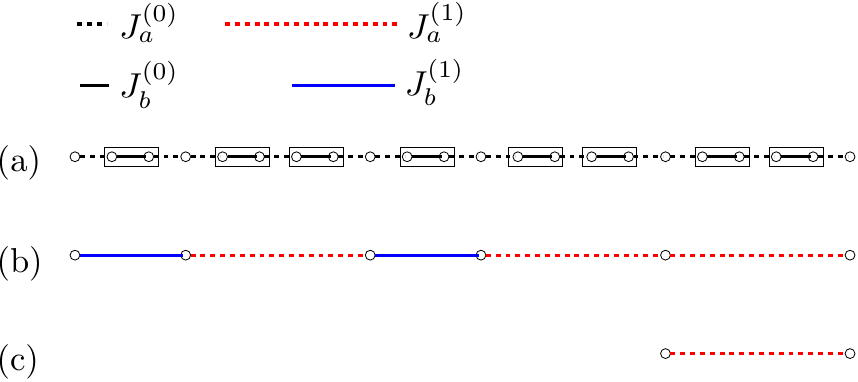} 
\caption{\label{fibonacci-chain-ap0-l22}
SDRG approach as applied to the spin-$1$ Fibonacci-Heisenberg chain
with $L=22$ sites. Sweeping over the effective chain in
(b) removes all effective $J_b$ bonds, leaving a single effective
$J_a$ bond, as shown in (c).}
\end{figure}
\begin{figure}
\includegraphics[width=0.99\columnwidth]{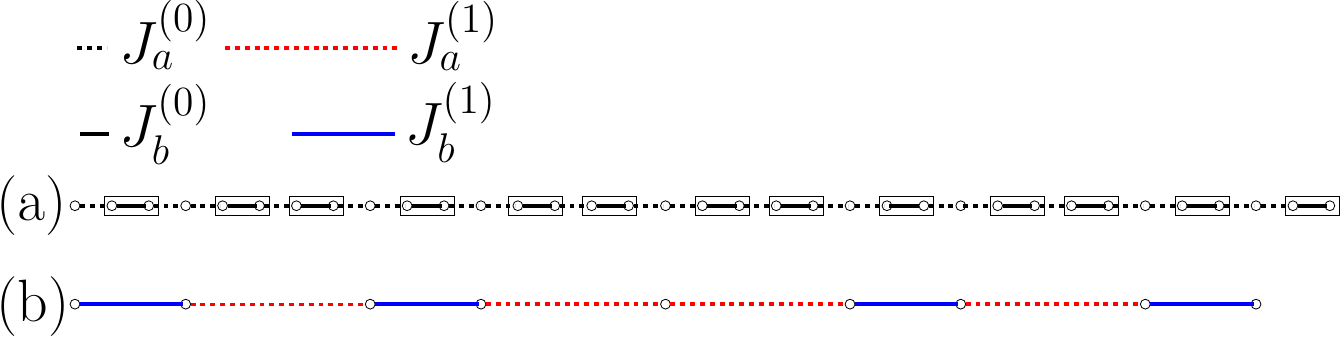} 
\caption{\label{fibonacci-chain-ap0-l35}
SDRG approach as applied to the spin-$1$ Fibonacci-Heisenberg chain
with $L=35$ sites. After sweeping over the effective chain in
(b), the central spin remains.}
\end{figure}

An SDRG estimate of the gap for finite open chains can be obtained
by stopping the RG scheme at the lowest energy scale for which
at least two spins are still active. Figures 
\ref{fibonacci-chain-ap0-l14} through \ref{fibonacci-chain-ap0-l35}
illustrate this for chains with $L=14$, $L=22$, and $L=35$ spins.
Since we want to obtain estimates to compare with the DMRG results
of Sec. \ref{sec:DMRG-Fibo}, we need to consider the gap between
the ground state, with total spin $S_T=0$ or $S_T=1$, and the lowest
energy level with $S_T=2$. For $L=14$ and $L=22$, the ground state
is a singlet ($S_T=0$), while for $L=35$, with an odd number of spins,
the ground state has total spin $S_T=1$.

As shown in Fig.~\ref{fibonacci-chain-ap0-l14}, for $L=14$ spins,
the lowest $S_T=2$ energy level corresponds to exciting two pairs
of spins to the local $S=1$ states. Thus, the SDRG estimate for
this gap is $2J_b^{(1)}$, with the effective $J_b^{(1)}$ calculated
from Eq. \eqref{eq:fibap0rec}. For $L=22$, as depicted in Fig.
~\ref{fibonacci-chain-ap0-l22}, the lowest $S_T=2$ excitation
comes from a local $S=2$ excitation of a pair of spins connected
by an effective $J_a^{(1)}$ bond, yielding a gap of $3J_a^{(1)}$.
Finally, since the ground state for $L=35$ has total spin
$S_T=1$, the lowest $S_T=2$ excitation corresponds to the local
excitation of a single spin pair, leading to a gap of $J_b^{(1)}$,
as shown in Fig. \ref{fibonacci-chain-ap0-l35}. Notice that
this explains the nonmonotonic behavior of the gaps with increasing 
system size, for $r\gtrsim 5$, as observed in Fig. \ref{fig:gap-fibo}.

Estimates of the SDRG gap for larger values of $L$ can be obtained
in an analogous fashion.

\bibliographystyle{apsrev4-1}
\bibliography{paper.bib}

%Merlin.mbs v4.21 2009-07-09.
\begin{thebibliography}{10}%
\makeatletter
\providecommand \@ifxundefined [1]{%
 \ifx #1\undefined \expandafter \@firstoftwo
 \else \expandafter \@secondoftwo
\fi
}%
\providecommand \@ifnum [1]{%
 \ifnum #1\expandafter \@firstoftwo
 \else \expandafter \@secondoftwo
\fi
}%
\providecommand \enquote [1]{``#1''}%
\providecommand \bibnamefont  [1]{#1}%
\providecommand \bibfnamefont [1]{#1}%
\providecommand \citenamefont [1]{#1}%
\providecommand\href[0]{\@sanitize\@href}%
\providecommand\@href[1]{\endgroup\@@startlink{#1}\endgroup\@@href}%
\providecommand\@@href[1]{#1\@@endlink}%
\providecommand \@sanitize [0]{\begingroup\catcode`\&12\catcode`\#12\relax}%
\@ifxundefined \pdfoutput {\@firstoftwo}{%
 \@ifnum{\z@=\pdfoutput}{\@firstoftwo}{\@secondoftwo}%
}{%
 \providecommand\@@startlink[1]{\leavevmode\special{html:<a href="#1">}}%
 \providecommand\@@endlink[0]{\special{html:</a>}}%
}{%
 \providecommand\@@startlink[1]{%
  \leavevmode
  \pdfstartlink
   attr{/Border[0 0 1 ]/H/I/C[0 1 1]}%
   user{/Subtype/Link/A<</Type/Action/S/URI/URI(#1)>>}%
  \relax
 }%
 \providecommand\@@endlink[0]{\pdfendlink}%
}%
\providecommand \url  [0]{\begingroup\@sanitize \@url }%
\providecommand \@url [1]{\endgroup\@href {#1}{\urlprefix}}%
\providecommand \urlprefix [0]{URL }%
\providecommand \Eprint[0]{\href }%
\@ifxundefined \urlstyle {%
  \providecommand \doi [1]{doi:\discretionary{}{}{}#1}%
}{%
  \providecommand \doi [0]{doi:\discretionary{}{}{}\begingroup
  \urlstyle{rm}\Url }%
}%
\providecommand \doibase [0]{http://dx.doi.org/}%
\providecommand \Doi[1]{\href{\doibase#1}}%
\providecommand \bibAnnote [3]{%
  \BibitemShut{#1}%
  \begin{quotation}\noindent
    \textsc{Key:}\ #2\\\textsc{Annotation:}\ #3%
  \end{quotation}%
}%
\providecommand \bibAnnoteFile [2]{%
  \IfFileExists{#2}{\bibAnnote {#1} {#2} {\input{#2}}}{}%
}%
\providecommand \typeout [0]{\immediate \write \m@ne }%
\providecommand \selectlanguage [0]{\@gobble}%
\providecommand \bibinfo [0]{\@secondoftwo}%
\providecommand \bibfield [0]{\@secondoftwo}%
\providecommand \translation [1]{[#1]}%
\providecommand \BibitemOpen[0]{}%
\providecommand \bibitemStop [0]{}%
\providecommand \bibitemNoStop [0]{.\EOS\space}%
\providecommand \EOS [0]{\spacefactor3000\relax}%
\providecommand \BibitemShut [1]{\csname bibitem#1\endcsname}%
%</preamble>
\bibitem{haldane83}%
  \BibitemOpen
  \bibfield{author}{%
  \bibinfo {author} {\bibfnamefont{F.~D.~M.}\ \bibnamefont{Haldane}},\ }%
  \bibfield{journal}{%
  \Doi{10.1103/PhysRevLett.50.1153}{\bibinfo {journal} {Phys. Rev. Lett.}}\ }%
  \textbf{\bibinfo {volume} {50}},\ \bibinfo {pages} {1153} (\bibinfo {year}
  {1983})%
  \bibAnnoteFile{NoStop}{haldane83}%
\bibitem{affleck89}%
  \BibitemOpen
  \bibfield{author}{%
  \bibinfo {author} {\bibfnamefont{I.}~\bibnamefont{Affleck}},\ }%
  \bibfield{journal}{%
  \Doi{10.1088/0953-8984/1/19/001}{\bibinfo {journal} {J. Phys.: Condens.
  Matter}}\ }%
  \textbf{\bibinfo {volume} {1}},\ \bibinfo {pages} {3047} (\bibinfo {year}
  {1989})%
  \bibAnnoteFile{NoStop}{affleck89}%
\bibitem{affleck87}%
  \BibitemOpen
  \bibfield{author}{%
  \bibinfo {author} {\bibfnamefont{I.}~\bibnamefont{Affleck}}, \bibinfo
  {author} {\bibfnamefont{T.}~\bibnamefont{Kennedy}}, \bibinfo {author}
  {\bibfnamefont{E.~H.}\ \bibnamefont{Lieb}},\ and\ \bibinfo {author}
  {\bibfnamefont{H.}~\bibnamefont{Tasaki}},\ }%
  \bibfield{journal}{%
  \Doi{10.1103/PhysRevLett.59.799}{\bibinfo {journal} {Phys. Rev. Lett.}}\ }%
  \textbf{\bibinfo {volume} {59}},\ \bibinfo {pages} {799} (\bibinfo {year}
  {1987})%
  \bibAnnoteFile{NoStop}{affleck87}%
\bibitem{dennijs}%
  \BibitemOpen
  \bibfield{author}{%
  \bibinfo {author} {\bibfnamefont{M.}~\bibnamefont{den Nijs}}\ and\ \bibinfo
  {author} {\bibfnamefont{K.}~\bibnamefont{Rommelse}},\ }%
  \bibfield{journal}{%
  \Doi{10.1103/PhysRevB.40.4709}{\bibinfo {journal} {Phys. Rev. B}}\ }%
  \textbf{\bibinfo {volume} {40}},\ \bibinfo {pages} {4709} (\bibinfo {month}
  {Sep}\ \bibinfo {year} {1989})%
  \bibAnnoteFile{NoStop}{dennijs}%
\bibitem{ma79}%
  \BibitemOpen
  \bibfield{author}{%
  \bibinfo {author} {\bibfnamefont{S.-K.}\ \bibnamefont{Ma}}, \bibinfo {author}
  {\bibfnamefont{C.}~\bibnamefont{Dasgupta}},\ and\ \bibinfo {author}
  {\bibfnamefont{C.-K.}\ \bibnamefont{Hu}},\ }%
  \bibfield{journal}{%
  \Doi{10.1103/PhysRevLett.43.1434}{\bibinfo {journal} {Phys. Rev. Lett.}}\ }%
  \textbf{\bibinfo {volume} {43}},\ \bibinfo {pages} {1434} (\bibinfo {year}
  {1979})%
  \bibAnnoteFile{NoStop}{ma79}%
\bibitem{dasgupta80}%
  \BibitemOpen
  \bibfield{author}{%
  \bibinfo {author} {\bibfnamefont{C.}~\bibnamefont{Dasgupta}}\ and\ \bibinfo
  {author} {\bibfnamefont{S.-K.}\ \bibnamefont{Ma}},\ }%
  \bibfield{journal}{%
  \Doi{10.1103/PhysRevB.22.1305}{\bibinfo {journal} {Phys. Rev. B}}\ }%
  \textbf{\bibinfo {volume} {22}},\ \bibinfo {pages} {1305} (\bibinfo {year}
  {1980})%
  \bibAnnoteFile{NoStop}{dasgupta80}%
\bibitem{doty92}%
  \BibitemOpen
  \bibfield{author}{%
  \bibinfo {author} {\bibfnamefont{C.~A.}\ \bibnamefont{Doty}}\ and\ \bibinfo
  {author} {\bibfnamefont{D.~S.}\ \bibnamefont{Fisher}},\ }%
  \bibfield{journal}{%
  \Doi{10.1103/PhysRevB.45.2167}{\bibinfo {journal} {Phys. Rev. B}}\ }%
  \textbf{\bibinfo {volume} {45}},\ \bibinfo {pages} {2167} (\bibinfo {year}
  {1992})%
  \bibAnnoteFile{NoStop}{doty92}%
\bibitem{fisher94}%
  \BibitemOpen
  \bibfield{author}{%
  \bibinfo {author} {\bibfnamefont{D.~S.}\ \bibnamefont{Fisher}},\ }%
  \bibfield{journal}{%
  \Doi{10.1103/PhysRevB.50.3799}{\bibinfo {journal} {Phys. Rev. B}}\ }%
  \textbf{\bibinfo {volume} {50}},\ \bibinfo {pages} {3799} (\bibinfo {year}
  {1994})%
  \bibAnnoteFile{NoStop}{fisher94}%
\bibitem{henelius98}%
  \BibitemOpen
  \bibfield{author}{%
  \bibinfo {author} {\bibfnamefont{P.}~\bibnamefont{Henelius}}\ and\ \bibinfo
  {author} {\bibfnamefont{S.~M.}\ \bibnamefont{Girvin}},\ }%
  \bibfield{journal}{%
  \Doi{10.1103/PhysRevB.57.11457}{\bibinfo {journal} {Phys. Rev. B}}\ }%
  \textbf{\bibinfo {volume} {57}},\ \bibinfo {pages} {11457} (\bibinfo {year}
  {1998})%
  \bibAnnoteFile{NoStop}{henelius98}%
\bibitem{Laflo04}%
  \BibitemOpen
  \bibfield{author}{%
  \bibinfo {author} {\bibfnamefont{N.}~\bibnamefont{Laflorencie}}, \bibinfo
  {author} {\bibfnamefont{H.}~\bibnamefont{Rieger}}, \bibinfo {author}
  {\bibfnamefont{A.~W.}\ \bibnamefont{Sandvik}},\ and\ \bibinfo {author}
  {\bibfnamefont{P.}~\bibnamefont{Henelius}},\ }%
  \bibfield{journal}{%
  \Doi{10.1103/PhysRevB.70.054430}{\bibinfo {journal} {Phys. Rev. B}}\ }%
  \textbf{\bibinfo {volume} {70}},\ \bibinfo {pages} {054430} (\bibinfo {year}
  {2004})%
  \bibAnnoteFile{NoStop}{Laflo04}%
\bibitem{hoyos07}%
  \BibitemOpen
  \bibfield{author}{%
  \bibinfo {author} {\bibfnamefont{J.~A.}\ \bibnamefont{Hoyos}}, \bibinfo
  {author} {\bibfnamefont{A.~P.}\ \bibnamefont{Vieira}}, \bibinfo {author}
  {\bibfnamefont{N.}~\bibnamefont{Laflorencie}},\ and\ \bibinfo {author}
  {\bibfnamefont{E.}~\bibnamefont{Miranda}},\ }%
  \bibfield{journal}{%
  \Doi{10.1103/PhysRevB.76.174425}{\bibinfo {journal} {Phys. Rev. B}}\ }%
  \textbf{\bibinfo {volume} {76}},\ \bibinfo {pages} {174425} (\bibinfo {year}
  {2007})%
  \bibAnnoteFile{NoStop}{hoyos07}%
\bibitem{damle02}%
  \BibitemOpen
  \bibfield{author}{%
  \bibinfo {author} {\bibfnamefont{K.}~\bibnamefont{Damle}},\ }%
  \bibfield{journal}{%
  \Doi{10.1103/PhysRevB.66.104425}{\bibinfo {journal} {Phys. Rev. B}}\ }%
  \textbf{\bibinfo {volume} {66}},\ \bibinfo {pages} {104425} (\bibinfo {year}
  {2002})%
  \bibAnnoteFile{NoStop}{damle02}%
\bibitem{saguia02}%
  \BibitemOpen
  \bibfield{author}{%
  \bibinfo {author} {\bibfnamefont{A.}~\bibnamefont{Saguia}}, \bibinfo {author}
  {\bibfnamefont{B.}~\bibnamefont{Boechat}},\ and\ \bibinfo {author}
  {\bibfnamefont{M.~A.}\ \bibnamefont{Continentino}},\ }%
  \bibfield{journal}{%
  \Doi{10.1103/PhysRevLett.89.117202}{\bibinfo {journal} {Phys. Rev. Lett.}}\
  }%
  \textbf{\bibinfo {volume} {89}},\ \bibinfo {pages} {117202} (\bibinfo {year}
  {2002})%
  \bibAnnoteFile{NoStop}{saguia02}%
\bibitem{hyman97}%
  \BibitemOpen
  \bibfield{author}{%
  \bibinfo {author} {\bibfnamefont{R.~A.}\ \bibnamefont{Hyman}}\ and\ \bibinfo
  {author} {\bibfnamefont{K.}~\bibnamefont{Yang}},\ }%
  \bibfield{journal}{%
  \Doi{10.1103/PhysRevLett.78.1783}{\bibinfo {journal} {Phys. Rev. Lett.}}\ }%
  \textbf{\bibinfo {volume} {78}},\ \bibinfo {pages} {1783} (\bibinfo {year}
  {1997})%
  \bibAnnoteFile{NoStop}{hyman97}%
\bibitem{hyman96phd}%
  \BibitemOpen
  \bibfield{author}{%
  \bibinfo {author} {\bibfnamefont{R.~A.}\ \bibnamefont{Hyman}},\ }%
  Ph.D. thesis,\ \bibinfo {school} {Indiana University} (\bibinfo {year}
  {1996})%
  \bibAnnoteFile{NoStop}{hyman96phd}%
\bibitem{monthus97}%
  \BibitemOpen
  \bibfield{author}{%
  \bibinfo {author} {\bibfnamefont{C.}~\bibnamefont{Monthus}}, \bibinfo
  {author} {\bibfnamefont{O.}~\bibnamefont{Golinelli}},\ and\ \bibinfo {author}
  {\bibfnamefont{T.}~\bibnamefont{Jolicoeur}},\ }%
  \bibfield{journal}{%
  \Doi{10.1103/PhysRevLett.79.3254}{\bibinfo {journal} {Phys. Rev. Lett.}}\ }%
  \textbf{\bibinfo {volume} {79}},\ \bibinfo {pages} {3254} (\bibinfo {year}
  {1997})%
  \bibAnnoteFile{NoStop}{monthus97}%
\bibitem{monthus98}%
  \BibitemOpen
  \bibfield{author}{%
  \bibinfo {author} {\bibfnamefont{C.}~\bibnamefont{Monthus}}, \bibinfo
  {author} {\bibfnamefont{O.}~\bibnamefont{Golinelli}},\ and\ \bibinfo {author}
  {\bibfnamefont{T.}~\bibnamefont{Jolicoeur}},\ }%
  \bibfield{journal}{%
  \Doi{10.1103/PhysRevB.58.805}{\bibinfo {journal} {Phys. Rev. B}}\ }%
  \textbf{\bibinfo {volume} {58}},\ \bibinfo {pages} {805} (\bibinfo {year}
  {1998})%
  \bibAnnoteFile{NoStop}{monthus98}%
\bibitem{bergkvist02}%
  \BibitemOpen
  \bibfield{author}{%
  \bibinfo {author} {\bibfnamefont{S.}~\bibnamefont{Bergkvist}}, \bibinfo
  {author} {\bibfnamefont{P.}~\bibnamefont{Henelius}},\ and\ \bibinfo {author}
  {\bibfnamefont{A.}~\bibnamefont{Rosengren}},\ }%
  \bibfield{journal}{%
  \Doi{10.1103/PhysRevB.66.134407}{\bibinfo {journal} {Phys. Rev. B}}\ }%
  \textbf{\bibinfo {volume} {66}},\ \bibinfo {pages} {134407} (\bibinfo {year}
  {2002})%
  \bibAnnoteFile{NoStop}{bergkvist02}%
\bibitem{lajko05}%
  \BibitemOpen
  \bibfield{author}{%
  \bibinfo {author} {\bibfnamefont{P.}~\bibnamefont{Lajk\'o}}, \bibinfo
  {author} {\bibfnamefont{E.}~\bibnamefont{Carlon}}, \bibinfo {author}
  {\bibfnamefont{H.}~\bibnamefont{Rieger}},\ and\ \bibinfo {author}
  {\bibfnamefont{F.}~\bibnamefont{Igl\'oi}},\ }%
  \bibfield{journal}{%
  \Doi{10.1103/PhysRevB.72.094205}{\bibinfo {journal} {Phys. Rev. B}}\ }%
  \textbf{\bibinfo {volume} {72}},\ \bibinfo {pages} {094205} (\bibinfo {year}
  {2005})%
  \bibAnnoteFile{NoStop}{lajko05}%
\bibitem{nishiyama1998}%
  \BibitemOpen
  \bibfield{author}{%
  \bibinfo {author} {\bibfnamefont{Y.}~\bibnamefont{Nishiyama}},\ }%
  \bibfield{journal}{%
  \Doi{10.1016/S0378-4371(97)00616-X}{\bibinfo {journal} {Physica A}}\ }%
  \textbf{\bibinfo {volume} {252}},\ \bibinfo {pages} {35} (\bibinfo {year}
  {1998})%
  \bibAnnoteFile{NoStop}{nishiyama1998}%
\bibitem{hida1999}%
  \BibitemOpen
  \bibfield{author}{%
  \bibinfo {author} {\bibfnamefont{K.}~\bibnamefont{Hida}},\ }%
  \bibfield{journal}{%
  \Doi{10.1103/PhysRevLett.83.3297}{\bibinfo {journal} {Phys. Rev. Lett.}}\ }%
  \textbf{\bibinfo {volume} {83}},\ \bibinfo {pages} {3297} (\bibinfo {year}
  {1999})%
  \bibAnnoteFile{NoStop}{hida1999}%
\bibitem{yang2000}%
  \BibitemOpen
  \bibfield{author}{%
  \bibinfo {author} {\bibfnamefont{K.}~\bibnamefont{Yang}}\ and\ \bibinfo
  {author} {\bibfnamefont{R.~A.}\ \bibnamefont{Hyman}},\ }%
  \bibfield{journal}{%
  \Doi{10.1103/PhysRevLett.84.2044}{\bibinfo {journal} {Phys. Rev. Lett.}}\ }%
  \textbf{\bibinfo {volume} {84}},\ \bibinfo {pages} {2044} (\bibinfo {year}
  {2000})%
  \bibAnnoteFile{NoStop}{yang2000}%
\bibitem{todo2000}%
  \BibitemOpen
  \bibfield{author}{%
  \bibinfo {author} {\bibfnamefont{S.}~\bibnamefont{Todo}}, \bibinfo {author}
  {\bibfnamefont{K.}~\bibnamefont{Kato}},\ and\ \bibinfo {author}
  {\bibfnamefont{H.}~\bibnamefont{Takayama}},\ }%
  \bibfield{journal}{%
  \bibinfo {journal} {J. Phys. Soc. Japan}\ }%
  \textbf{\bibinfo {volume} {69A}},\ \bibinfo {pages} {355} (\bibinfo {year}
  {2000})%
  \bibAnnoteFile{NoStop}{todo2000}%
\bibitem{griffiths69}%
  \BibitemOpen
  \bibfield{author}{%
  \bibinfo {author} {\bibfnamefont{R.~B.}\ \bibnamefont{Griffiths}},\ }%
  \bibfield{journal}{%
  \Doi{10.1103/PhysRevLett.23.17}{\bibinfo {journal} {Phys. Rev. Lett.}}\ }%
  \textbf{\bibinfo {volume} {23}},\ \bibinfo {pages} {17} (\bibinfo {year}
  {1969})%
  \bibAnnoteFile{NoStop}{griffiths69}%
\bibitem{schechtman84}%
  \BibitemOpen
  \bibfield{author}{%
  \bibinfo {author} {\bibfnamefont{D.}~\bibnamefont{Schechtman}}, \bibinfo
  {author} {\bibfnamefont{I.}~\bibnamefont{Blech}}, \bibinfo {author}
  {\bibfnamefont{D.}~\bibnamefont{Gratias}},\ and\ \bibinfo {author}
  {\bibfnamefont{J.~W.}\ \bibnamefont{Cahn}},\ }%
  \bibfield{journal}{%
  \Doi{10.1103/PhysRevLett.53.1951}{\bibinfo {journal} {Phys. Rev. Lett.}}\ }%
  \textbf{\bibinfo {volume} {53}},\ \bibinfo {pages} {1951} (\bibinfo {year}
  {1984})%
  \bibAnnoteFile{NoStop}{schechtman84}%
\bibitem{vieira05a}%
  \BibitemOpen
  \bibfield{author}{%
  \bibinfo {author} {\bibfnamefont{A.~P.}\ \bibnamefont{Vieira}},\ }%
  \bibfield{journal}{%
  \Doi{10.1103/PhysRevLett.94.077201}{\bibinfo {journal} {Phys. Rev. Lett.}}\
  }%
  \textbf{\bibinfo {volume} {94}},\ \bibinfo {pages} {077201} (\bibinfo {year}
  {2005})%
  \bibAnnoteFile{NoStop}{vieira05a}%
\bibitem{vieira05b}%
  \BibitemOpen
  \bibfield{author}{%
  \bibinfo {author} {\bibfnamefont{A.~P.}\ \bibnamefont{Vieira}},\ }%
  \bibfield{journal}{%
  \Doi{10.1103/PhysRevB.71.134408}{\bibinfo {journal} {Phys. Rev. B}}\ }%
  \textbf{\bibinfo {volume} {71}},\ \bibinfo {pages} {134408} (\bibinfo {year}
  {2005})%
  \bibAnnoteFile{NoStop}{vieira05b}%
\bibitem{igloi05}%
  \BibitemOpen
  \bibfield{author}{%
  \bibinfo {author} {\bibfnamefont{F.}~\bibnamefont{Igl\'oi}}\ and\ \bibinfo
  {author} {\bibfnamefont{C.}~\bibnamefont{Monthus}},\ }%
  \bibfield{journal}{%
  \Doi{10.1016/j.physrep.2005.02.006}{\bibinfo {journal} {Phys. Rep.}}\ }%
  \textbf{\bibinfo {volume} {412}},\ \bibinfo {pages} {277} (\bibinfo {year}
  {2005})%
  \bibAnnoteFile{NoStop}{igloi05}%
\bibitem{hida92}%
  \BibitemOpen
  \bibfield{author}{%
  \bibinfo {author} {\bibfnamefont{K.}~\bibnamefont{Hida}},\ }%
  \bibfield{journal}{%
  \Doi{10.1103/PhysRevB.45.2207}{\bibinfo {journal} {Phys. Rev. B}}\ }%
  \textbf{\bibinfo {volume} {45}},\ \bibinfo {pages} {2207} (\bibinfo {year}
  {1992})%
  \bibAnnoteFile{NoStop}{hida92}%
\bibitem{Sandvik91}%
  \BibitemOpen
  \bibfield{author}{%
  \bibinfo {author} {\bibfnamefont{A.}~\bibnamefont{Sandvik}}\ and\ \bibinfo
  {author} {\bibfnamefont{J.}~\bibnamefont{Kurkij\"arvi}},\ }%
  \bibfield{journal}{%
  \Doi{10.1103/PhysRevB.43.5950}{\bibinfo {journal} {Phys. Rev. B}}\ }%
  \textbf{\bibinfo {volume} {43}},\ \bibinfo {pages} {5950} (\bibinfo {year}
  {1991})%
  \bibAnnoteFile{NoStop}{Sandvik91}%
\bibitem{Sandvik92}%
  \BibitemOpen
  \bibfield{author}{%
  \bibinfo {author} {\bibfnamefont{A.}~\bibnamefont{Sandvik}},\ }%
  \bibfield{journal}{%
  \Doi{10.1088/0305-4470/25/13/017}{\bibinfo {journal} {J. Phys A.}}\ }%
  \textbf{\bibinfo {volume} {25}},\ \bibinfo {pages} {3667} (\bibinfo {year}
  {1992})%
  \bibAnnoteFile{NoStop}{Sandvik92}%
\bibitem{Syljuasen02}%
  \BibitemOpen
  \bibfield{author}{%
  \bibinfo {author} {\bibfnamefont{O.~F.}\ \bibnamefont{Sylju\aa{}sen}}\ and\
  \bibinfo {author} {\bibfnamefont{A.}~\bibnamefont{Sandvik}},\ }%
  \bibfield{journal}{%
  \Doi{10.1103/PhysRevE.66.046701}{\bibinfo {journal} {Phys. Rev. E}}\ }%
  \textbf{\bibinfo {volume} {66}},\ \bibinfo {pages} {046701} (\bibinfo {year}
  {2002})%
  \bibAnnoteFile{NoStop}{Syljuasen02}%
\bibitem{white92}%
  \BibitemOpen
  \bibfield{author}{%
  \bibinfo {author} {\bibfnamefont{S.~R.}\ \bibnamefont{White}},\ }%
  \bibfield{journal}{%
  \Doi{10.1103/PhysRevLett.69.2863}{\bibinfo {journal} {Phys. Rev. Lett.}}\ }%
  \textbf{\bibinfo {volume} {69}},\ \bibinfo {pages} {2863} (\bibinfo {year}
  {1992})%
  \bibAnnoteFile{NoStop}{white92}%
\bibitem{McCulloch07}%
  \BibitemOpen
  \bibfield{author}{%
  \bibinfo {author} {\bibfnamefont{I.~P.}\ \bibnamefont{McCulloch}},\ }%
  \bibfield{journal}{%
  \Doi{10.1088/1742-5468/2007/10/P10014}{\bibinfo {journal} {J. Stat. Mech}},\
  \bibinfo {pages} {P10014}}%
   (\bibinfo {year} {2007})%
  \bibAnnoteFile{NoStop}{McCulloch07}%
\bibitem{McCulloch02}%
  \BibitemOpen
  \bibfield{author}{%
  \bibinfo {author} {\bibfnamefont{I.~P.}\ \bibnamefont{McCulloch}}\ and\
  \bibinfo {author} {\bibfnamefont{M.}~\bibnamefont{Gul\'acsi}},\ }%
  \bibfield{journal}{%
  \Doi{10.1209/epl/i2002-00393-0}{\bibinfo {journal} {Europhys. Lett.}}\ }%
  \textbf{\bibinfo {volume} {57}},\ \bibinfo {pages} {852} (\bibinfo {year}
  {2002})%
  \bibAnnoteFile{NoStop}{McCulloch02}%
\bibitem{Note1}%
  \BibitemOpen
  \bibinfo {note} {We also performed calculations for periodic boundary
  conditions, but convergence was far more problematic than for open chains.}%
  \bibAnnoteFile{Stop}{Note1}%
\bibitem{Note2}%
  \BibitemOpen
  \bibinfo {note} {Of course, for weak or moderate modulation further-neighbor
  couplings extend over larger and larger distances, making numerical
  implementations of the SDRG quite intricate. These cases are not considered
  here.}%
  \bibAnnoteFile{Stop}{Note2}%
\bibitem{oliveira2012}%
  \BibitemOpen
  \bibfield{author}{%
  \bibinfo {author} {\bibfnamefont{F.~J.}\ \bibnamefont{{Oliveira Filho}}},
  \bibinfo {author} {\bibfnamefont{M.~S.}\ \bibnamefont{Faria}},\ and\ \bibinfo
  {author} {\bibfnamefont{A.~P.}\ \bibnamefont{Vieira}},\ }%
  \bibfield{journal}{%
  \Doi{10.1088/1742-5468/2012/03/P03007}{\bibinfo {journal} {J. Stat. Mech.}}\
  }%
  \textbf{\bibinfo {volume} {2012}},\ \bibinfo {pages} {P03007} (\bibinfo
  {year} {2012})%
  \bibAnnoteFile{NoStop}{oliveira2012}%
\bibitem{barghathi2013}%
  \BibitemOpen
  \bibfield{author}{%
  \bibinfo {author} {\bibfnamefont{H.}~\bibnamefont{Barghathi}}, \bibinfo
  {author} {\bibfnamefont{D.}~\bibnamefont{Nozadze}},\ and\ \bibinfo {author}
  {\bibfnamefont{T.}~\bibnamefont{Vojta}},\ }%
  \bibfield{journal}{%
  \Doi{10.1103/PhysRevE.89.012112}{\bibinfo {journal} {Phys. Rev. E}}\ }%
  \textbf{\bibinfo {volume} {89}},\ \bibinfo {pages} {012112} (\bibinfo {year}
  {2014})%
  \bibAnnoteFile{NoStop}{barghathi2013}%
\bibitem{Note3}%
  \BibitemOpen
  \bibinfo {note} {See http://physics.uq.edu.au/people/ianmcc/mptoolkit/}%
  \bibAnnoteFile{NoStop}{Note3}%
\bibitem{Alet05}%
  \BibitemOpen
  \bibfield{author}{%
  \bibinfo {author} {\bibfnamefont{F.}~\bibnamefont{Alet}}, \bibinfo {author}
  {\bibfnamefont{S.}~\bibnamefont{Wessel}},\ and\ \bibinfo {author}
  {\bibfnamefont{M.}~\bibnamefont{Troyer}},\ }%
  \bibfield{journal}{%
  \Doi{10.1103/PhysRevE.71.036706}{\bibinfo {journal} {Phys. Rev. E}}\ }%
  \textbf{\bibinfo {volume} {71}},\ \bibinfo {pages} {036706} (\bibinfo {month}
  {Mar}\ \bibinfo {year} {2005})%
  \bibAnnoteFile{NoStop}{Alet05}%
\bibitem{ALPS}%
  \BibitemOpen
  \bibfield{author}{%
  \bibinfo {author} {\bibfnamefont{A.}~\bibnamefont{Albuquerque}}, \bibinfo
  {author} {\bibfnamefont{F.}~\bibnamefont{Alet}}, \bibinfo {author}
  {\bibfnamefont{P.}~\bibnamefont{Corboz}}, \bibinfo {author}
  {\bibfnamefont{P.}~\bibnamefont{Dayal}}, \bibinfo {author}
  {\bibfnamefont{A.}~\bibnamefont{Feiguin}}, \bibinfo {author}
  {\bibfnamefont{S.}~\bibnamefont{Fuchs}}, \bibinfo {author}
  {\bibfnamefont{L.}~\bibnamefont{Gamper}}, \bibinfo {author}
  {\bibfnamefont{E.}~\bibnamefont{Gull}}, \bibinfo {author}
  {\bibfnamefont{S.}~\bibnamefont{G\"urtler}}, \bibinfo {author}
  {\bibfnamefont{A.}~\bibnamefont{Honecker}}, \bibinfo {author}
  {\bibfnamefont{R.}~\bibnamefont{Igarashi}}, \bibinfo {author}
  {\bibfnamefont{M.}~\bibnamefont{K\"orner}}, \bibinfo {author}
  {\bibfnamefont{A.}~\bibnamefont{Kozhevnikov}}, \bibinfo {author}
  {\bibfnamefont{A.}~\bibnamefont{L{\"a}uchli}}, \bibinfo {author}
  {\bibfnamefont{S.}~\bibnamefont{Manmana}}, \bibinfo {author}
  {\bibfnamefont{M.}~\bibnamefont{Matsumoto}}, \bibinfo {author}
  {\bibfnamefont{I.}~\bibnamefont{McCulloch}}, \bibinfo {author}
  {\bibfnamefont{F.}~\bibnamefont{Michel}}, \bibinfo {author}
  {\bibfnamefont{R.}~\bibnamefont{Noack}}, \bibinfo {author}
  {\bibfnamefont{G.~P.}\ \bibnamefont{owski}}, \bibinfo {author}
  {\bibfnamefont{L.}~\bibnamefont{Pollet}}, \bibinfo {author}
  {\bibfnamefont{T.}~\bibnamefont{Pruschke}}, \bibinfo {author}
  {\bibfnamefont{U.}~\bibnamefont{Schollw{\"o}ck}}, \bibinfo {author}
  {\bibfnamefont{S.}~\bibnamefont{Todo}}, \bibinfo {author}
  {\bibfnamefont{S.}~\bibnamefont{Trebst}}, \bibinfo {author}
  {\bibfnamefont{M.}~\bibnamefont{Troyer}}, \bibinfo {author}
  {\bibfnamefont{P.}~\bibnamefont{Werner}},\ and\ \bibinfo {author}
  {\bibfnamefont{S.}~\bibnamefont{Wessel}},\ }%
  \bibfield{journal}{%
  \Doi{10.1016/j.jmmm.2006.10.304}{\bibinfo {journal} {J. Magn. Magn. Mater.}}\
  }%
  \textbf{\bibinfo {volume} {310}},\ \bibinfo {pages} {1187 } (\bibinfo {year}
  {2007}),\ ISSN \bibinfo {issn} {0304-8853}%
  \bibAnnoteFile{NoStop}{ALPS}%
\bibitem{Note4}%
  \BibitemOpen
  \bibinfo {note} {See http://alps.comp-phys.org}%
  \bibAnnoteFile{NoStop}{Note4}%
\end{thebibliography}%

\end{document}